\documentclass[twocolumn,superscriptaddress]{revtex4}

\usepackage{amsmath, amsfonts, amssymb, graphicx, color,xcolor}
\usepackage{ulem}
\graphicspath{{./figures_src/}}
\newcommand{\beq}{\begin{equation}}
	\newcommand{\eeq}{\end{equation}}
      \newcommand{\app}{{Appendix}}
            \newcommand{\SFig}[1]{Fig.~#1}
\newcommand{\quotes}[1]{``#1''}
\newcommand{\lastequal}{Corresponding authors. These authors contributed equally.}
\newcommand{\rev}[1]{{#1}}

\newcommand{\VC}{}

	\newcommand{\MM}[1]{{Sec.~\ref{#1}}}
	
\begin{document}
	
\title{Evolutionary stability of antigenically escaping viruses}

\author{Victor Chard\`es}
\altaffiliation{These authors contributed equally to this work}
\affiliation{Laboratoire de physique de l'\'Ecole normale sup\'erieure,
	CNRS, PSL University, Sorbonne Universit\'e, and Universit\'e 
	Paris-Cit\'e, 75005 Paris, France}
\affiliation{Center for Computational Biology, Flatiron Institute, New York, New York 10010, USA}
\author{Andrea Mazzolini}
\altaffiliation{These authors contributed equally to this work}
\affiliation{Laboratoire de physique de l'\'Ecole normale sup\'erieure,
	CNRS, PSL University, Sorbonne Universit\'e, and Universit\'e 
	Paris-Cit\'e, 75005 Paris, France}
\author{Thierry Mora}
\thanks{\lastequal}
\affiliation{Laboratoire de physique de l'\'Ecole normale sup\'erieure,
	CNRS, PSL University, Sorbonne Universit\'e, and Universit\'e 
	Paris-Cit\'e, 75005 Paris, France}
\author{Aleksandra M. Walczak}
\thanks{\lastequal}
\affiliation{Laboratoire de physique de l'\'Ecole normale sup\'erieure,
	CNRS, PSL University, Sorbonne Universit\'e, and Universit\'e 
	Paris-Cit\'e, 75005 Paris, France}
	
\begin{abstract}
Antigenic variation is the main immune escape mechanism for RNA viruses like influenza or SARS-CoV-2. While high mutation rates promote antigenic escape, they also induce large mutational loads and reduced fitness. It remains unclear how this cost-benefit trade-off selects the mutation rate of viruses. Using a traveling wave model for the co-evolution of viruses and host immune systems in a finite population, we investigate how immunity affects the evolution of the mutation rate and other non-antigenic traits, such as virulence.
We first show that the nature of the wave depends on how cross-reactive immune systems are, reconciling previous approaches. The immune-virus system behaves like a Fisher wave at low cross-reactivities, and like a fitness wave at high cross-reactivities.
These regimes predict different outcomes for the evolution of non-antigenic traits. At low cross-reactivities, the evolutionarily stable strategy is to maximize the speed of the wave, implying a higher mutation rate and increased virulence.
At large cross-reactivities, where our estimates place H3N2 influenza, the stable strategy is to increase the basic reproductive number, keeping the mutation rate to a minimum and virulence low.

\end{abstract}

\maketitle

\section{Introduction}

RNA viruses like influenza or SARS-Cov-2 are subject to a constant antigenic evolution driven by their hosts' immune pressure and fueled by their remarkably high mutation rates \cite{Belshaw2008, Sanjuan2010, Duffy2018, Peck2018}. 
Although immune memories in hosts are geared towards reinfections and possible variants \cite{Chardes2022, Shlomchik2019, Viant2020} this constant evolution allows viruses to evade immunity, leading to repeated epidemics and reinfections. 
The recent SARS-Cov-2 outbreak has shown that the management of infectious diseases remains a global health challenge and is now a major public concern in the face of an increased ecosystem disruption \cite{Jones2008, Jones2013, Salkeld2015}. 
In these conditions, predicting the emergence of future variants is essential to inform vaccine strain selection, improve collective immunity and lift the burden imposed on healthcare systems.

These challenges have led to the development of theoretical methods to predict influenza antigenic evolution \cite{Neher2016, Luksza2014, Morris2018}. 
However, these approaches do not inform about the evolution of non-antigenic traits like virulence or the mutation rate itself, {\VC while} these traits clearly influence the future state of the viral and host populations. 
 {\VC On the other hand}, extensive epidemiological literature describes host-pathogen co-evolution in pathogens not escaping immunity \cite{May1990, Anderson1992, Mideo2008, Alizon2009, Lion2018}. 
While bridges between this literature and population genetics models have long been built to predict the evolution of parasite virulence \cite{Day2007, Gandon2016}, they have only recently been extended to study virulence evolution in antigenically evolving viruses \cite{Sasaki2021}. 
These new approaches showed that in populations of infinite size, antigenic escape promotes higher transmission rates and virulence than expected for pathogens at an endemic equilibrium.
However, antigenic adaptation is mostly driven by stochastic birth, death and mutation events occurring in the most well adapted individuals \cite{Hallatschek2011, Desai2007}, which are typically in small numbers. 
Thus, finite size demographic effects are crucial to accurately describe antigenically evolving pathogens \cite{Rouzine2018, Tsimring1996, Minayev2009, Marchi2021}. 
It remains unclear how antigenic escape in a co-evolving system of viruses and antibodies, coupled with finite size demography, constrains the evolution of non-antigenic traits, such as the mutation rate or the virulence.

To model antigenic escape it is convenient to describe both the host immune memories and the viral strains as living in the same antigenic space \cite{Rouzine2018, Marchi2021, Sasaki2021, Gog2002}, corresponding to a space of molecular similarity, also called ``shape space'' \cite{Segel1989}. 
This construction is not just conceptual: dimensionality reduction of hemmaglutination inhibition data can be used to build low dimensional manifolds on which influenza and hosts antibodies co-evolve \cite{Smith2004, Bedford2014, Fonville2014}. 
While mapping influenza evolution in this shape space has been used to describe evolutionary modes of influenza \cite{Marchi2021, Yan2019, Bedford2012a}, it remains unclear how this regime influences the evolution of non-antigenic traits such as the mutation rate of viruses.

In this work, we describe with a SI(R) formalism \cite{Kermack1927, Anderson1992} for the co-evolution of a finite population of viruses and immune systems of infected hosts, in an effective one dimensional antigenic space. 
The model generates a traveling wave for the number of viruses, which escapes from a continuously adapting immune system.
Our first finding is that, depending on the model parameters, the antigenic wave crosses over between two well characterized regimes.
Among those parameters, the ability of antibodies to recognize pathogens similar to the already encountered ones, called immune cross-reactivity or cross-immunity, plays a key role.
A narrow cross-reactivity leads to a Fisher-Kolmogorov–Petrovsky–Piskunov (FKPP) traveling wave \cite{Fisher1937}, while, for a large one, the wave converges to a linear-fitness wave dominated by finite-size effects \cite{Tsimring1996, Cohen2005}, with a smooth crossover between the two.
These regimes result in very different scalings of the macroscopic properties of the wave, such as its speed.
They also affect the evolutionary stability of non-antigenic traits.
To investigate this effect, we derive a simple and general relation for the evolutionary stability of viral parameters, which displays two qualitatively different behaviors depending on the regimes.
We then apply these results to study the evolution of the viral mutation rate on the one hand, and of the virulence on the other. We discuss how the evolutionary stable states are strongly impacted by cross-reactivity.


\section{Results}
\subsection{Model of co-evolving pathogens and immune systems}

\begin{figure}
\includegraphics[width=.48\textwidth]{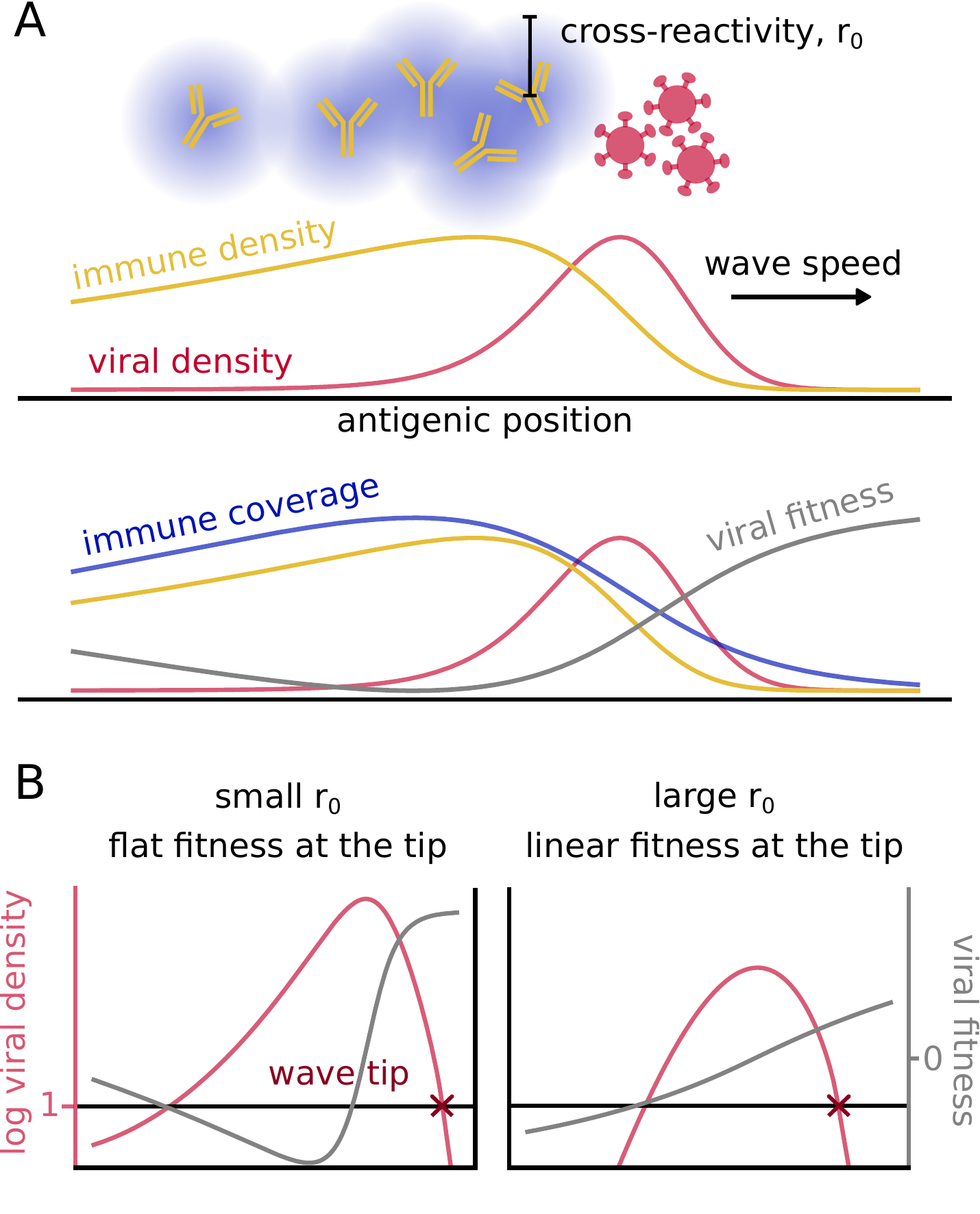}
\caption{{\bf Two types of antigenic waves controled by cross-reactivity.} (A): Typical traveling wave dynamics of a population of viruses (in red) escaping the immune system (in blue), equations \ref{eq:dyn_n}, \ref{eq:dyn_n_cutoff}.
In the bottom panel the immune coverage and the viral fitness are also
drawn. \rev{The increase of fitness in the back of the wave does not
  give rise to a second wave as the virus is extinct in that region.}
(B): Differences between the two regimes of small and large cross reactivity.
The plots show the viral density (in red) under two different fitness profiles (in yellow).
They highlight the phenomenological difference between the two regimes.}
\label{fig1}
\end{figure}

We start by defining and analyzing a mathematical
model of pathogen-immune dynamics.
We consider the evolution of a pathogen in a one-dimensional antigenic
space with density $n(x,t)$ (Fig.~\ref{fig1}A, red line), where $x$
denotes position in that space. Immune protections are also assigned positions
in that space, so that protections are close to the viruses they recognize.
While the antigenic space is believed to have higher dimensions
\cite{perelson1979theoretical}, theoretical work has shown that the effective evolution of the
resulting traveling wave of escaping viruses is ``canalized'' into a one-dimensional
track \cite{Bedford2012a}, provided that
we ignore possible speciation events \cite{Marchi2019,
  Yan2019,Marchi2021}. This picture is overall consistent with
influenza data that show a low-dimensional reduction
of the viral evolutionary trajectory from hemagglutination inhibition
assays \cite{Smith2004}.

We assume that mutations act continuously and in an unbiased way on
the antigenic space so that, in the limit of infinitely small
mutations happening at a rate $\mu_x$, the density of infected hosts
effectively diffuses with constant $D = \mu_x \Delta x^2/2$, where
$\Delta x$ is the typical step covered by a mutation in the antigenic
space.
\rev{This continuous model approximates any arbitrary discrete mutational model provided that a random amino acid mutation in the antigenic sites of the pathogen is unlikely to induce a large change in hemagglutination inhibition titer, or equivalently in the antigenicity of the strain. For influenza, in-spite of rare mutations inducing large antigenic jumps \cite{Smith2004, koel2013substitutions}, the typical number of amino acid mutations happening in the hemaglutinin antigenic sites before observing a substantial change in antigenicity is closer to 15 substitutions \cite{Luksza2014}. In this regime and for similar pathogens, a continuous approximation is accurate enough.}

Following standard SI(R) modeling, we denote by $\beta$ the pathogen
transmission rate in absence of immunity, $\alpha$ its virulence and $\gamma$ the recovery
rate.
An important quantity is the reproductive ratio, $R_0 = \beta/(\alpha
+ \gamma)$, corresponding to the mean number of transmissions
infectious individuals cause in an unprotected population, before they recover or die. We define the
effective growth rate of a viral strain at position $x$ as $F(x,t) =
\beta S(x,t) - \alpha - \gamma$, where $S(x,t)$ is the susceptibility
of the population to that strain, defined below.
This leads to the following stochastic differential equation for the viral evolution:
\begin{equation}
\begin{aligned}
\partial_t n(x,t) =& \;F(x,t)n(x,t) + D \partial^2_x n(x,t) + \\
& + \text{demographic noise}.
\label{eq:dyn_n}
\end{aligned}
\end{equation}

We consider an effective, population-averaged effect of the immune
systems of the hosts onto the virus \cite{Marchi2021}.
The immune protection is decribed by a function $h(x,t)$ (Fig.~\ref{fig1}A,
blue line), which is the probability density of
immune receptors in a random host.
We consider $N_h$ hosts, each with $M$ immune protections drawn at
random from $h(x,t)$.
Upon infection by $x$, the host acquires a new immune protection at
$x$, which replaces one of its $M$ protections at random. This results
in the following population-wide dynamics:
\begin{equation}
\partial_t h(x,t) = \frac{1}{M N_h} \left[ n(x,t) - N(t) h(x,t) \right], 
\label{eq:dyn_h}
\end{equation}
where $N(t) = \int n(x,t) dx $ is the total number of infected hosts.

To estimate the susceptibility $S(x,t)$, we assume that a protection at
position $x$ provides protection against nearby pathogens, with a
probability that decays exponentially with characteristic length $r_0$, called
cross-reactivity range. This allows us to define the immune coverage as
\begin{equation}
c(x,t) = \int dy\, h(y,t) e^{-\frac{|x - y|}{r_0}}, \label{eq:coverage}
\end{equation}
which is the probability that a random protection
from a random host is effective against $x$ (Fig.~\ref{fig1}A, green
curve).
The susceptibility is then just the probability that none of the
$M$ protections of a random host is effective against $x$:
\begin{equation}
S(x, t) = \left( 1 - c(x,t) \right)^M .
\label{eq:susceptibility}
\end{equation}

\subsection{Deterministic approximation}

Eqs.~\ref{eq:dyn_n}--\ref{eq:susceptibility} describe the
co-evolutionary dynamics of pathogens and immune protections.
Because escape mutants of the pathogen are subject to random
extinctions due to genetic drift in their early days, one should not ignore demographic noise for all population sizes.
To simplify the computational load yet account for the effect of small
numbers, we use a deterministic
version of Eq.~\ref{eq:dyn_n} where viruses stop spreading when $n(x,t) < n_c$:
\begin{equation}
\partial_t n(x,t) = F(x,t)\Theta(n - n_c)n(x,t) + D \partial^2_x n(x,t), 
\label{eq:dyn_n_cutoff}
\end{equation}
where $\Theta(x)=1$ if $x>0$ and $0$ otherwise. 
This approximation is known to provide traveling wave results in
excellent agreement with fully stochastic agent-based simulations in
various models of rapidly adapting populations \cite{Cohen2005,
  Hallatschek2011, Marchi2021}.
Details about simulations of this equation are given in \MM{sec:simulations}. 
We checked that our results are consistent with a full stochastic approach, \MM{sec:stoch_sim}.

Note that in general
$n$ has units of an inverse antigenic distance.
We will show that results depend very weakly on $n_c$ (\SFig{S1}).
Nevertheless, to fix the scale of $x$ in an interpretable way, we set $\Delta x=1$, so
that $n(x,t)$ roughly represents the average number of infected hosts in its
mutation class (i.e. within a bin of size $\Delta x$). We then set 
$n_c=1$, which corresponds to one individual per class. The
cross-reactivity parameter $r_0$ can be interpreted as the number of mutations that a virus needs to acquire to escape an immune protection.

\begin{table}
\centering
\caption{List of free parameters of the model}
\begin{tabular}{c|c||c|c}
	$\mu_x$ & Viral mutation rate & $\beta$ & Viral transmission rate \\
	\hline
	$\Delta x$ & Mutational step & $r_0$ & Receptor cross reactivity \\
	\hline
	$\alpha$ & Virulence & $N_h$ & Number of hosts \\
	\hline
	$\gamma$ & Recovery rate & $M$ & Protections per host \\
\end{tabular}
\label{tab:parameters}
\end{table}

\begin{table}
	\centering
	\caption{Main quantities of the model}
	\begin{tabular}{c|c}
		$n(x,t)$, Eq.\ref{eq:dyn_n} & Density of infected hosts \\
		\hline
		$h(x,t)$, Eq.\ref{eq:dyn_h} & Density of immune receptors \\
		\hline
		$c(x,t)$, Eq.\ref{eq:coverage} & Coverage of the receptors \\
		\hline
		$S(x,t)$, Eq.\ref{eq:susceptibility} & Susceptibility \\
		\hline
		$F(x,t)=\beta S(x,t) -\alpha-\gamma$ & Viral growth rate \\		\hline
		$F_{\rm max} =\beta -\alpha-\gamma$ & Maximal viral growth
                                                  rate\\		\hline
		$D=\mu_x \Delta x^2/2$ & Mutation diffusion coef.\\
		\hline
		$R_0=\beta /(\alpha+\gamma)$ & Reproductive ratio\\
		\hline
		$v$ & Speed of the viral wave \\
		\hline
		$F_T=F(x_T)$ & Viral growth rate at the wave tip \\
		\hline
		$s_T=\partial_x F(x_T)$ & Slope of growth rate at the tip \\
		\hline
		$\sigma_T = \xi_0 ( D s_T^2 )^{1/3}$ & Notation  shorthand \\
          \hline
          $k=r_0^2F_{\rm max}/D$ & growth-to-escape dimensionless ratio
	\end{tabular}
	\label{tab:quantities}
\end{table}


\subsection{Cross-reactivity drives different regimes of antigenic evolution}
\label{sec:wave}

The coupled system of equations \ref{eq:dyn_n} and \ref{eq:dyn_h} admits a traveling wave solution, where the viral population is a moving bump followed by the immune system (Fig.\ref{fig1}A).
This dynamics is known to be driven mainly by the few individuals the front of the wave \cite{Desai2007, Hallatschek2011}: mutations generate new strains at more favorable antigenic positions ahead of the wave, where the hosts' immune systems provides less protection.
As a consequence, they grow faster than strains in the bulk of the wave, which is under stronger immune pressure.
This process is controlled by the few individuals at the front tip and therefore is intrinsically stochastic.

There are two different limits depending on the shape of the viral fitness at the tip of the wave, as illustrated in Fig.~\ref{fig1}B.
For small cross reactivities $r_0$ of the immune protections, viral strains at the tip feel no immune pressure at all. The fitness profile is thus locally constant.
As we will see in detail, the wave dynamics in this regime corresponds to the classical FKPP traveling wave \cite{Fisher1937}, where stochastic effects do not play an important role.

At large $r_0$, the immune coverage extends all the way to the tip of the wave, where viral strains experience a local gradient of fitness.
This case, where stochastic events at the wave tip drive its motion, has been also well characterized in previous works \cite{Tsimring1996,Cohen2005}.
In general, varying $r_0$ allows us to interpolate smoothly between the two regimes.

These two different behaviours are not just of technical interest. 
As we will see, they result in different parameter dependencies for the speed of the wave, and imply markedly different evolutionary stable states for non-antigenic traits such as the mutation rate or the virulence.
While previous work on the evolutionary stability of such traits has focused on the FKPP regime \cite{Sasaki2021}, here we treat the general case.

\subsection{The crossover results in different wave speed dependencies}
\label{sec:wave_speed}

\begin{figure}[th!]
	\includegraphics[width=.48\textwidth]{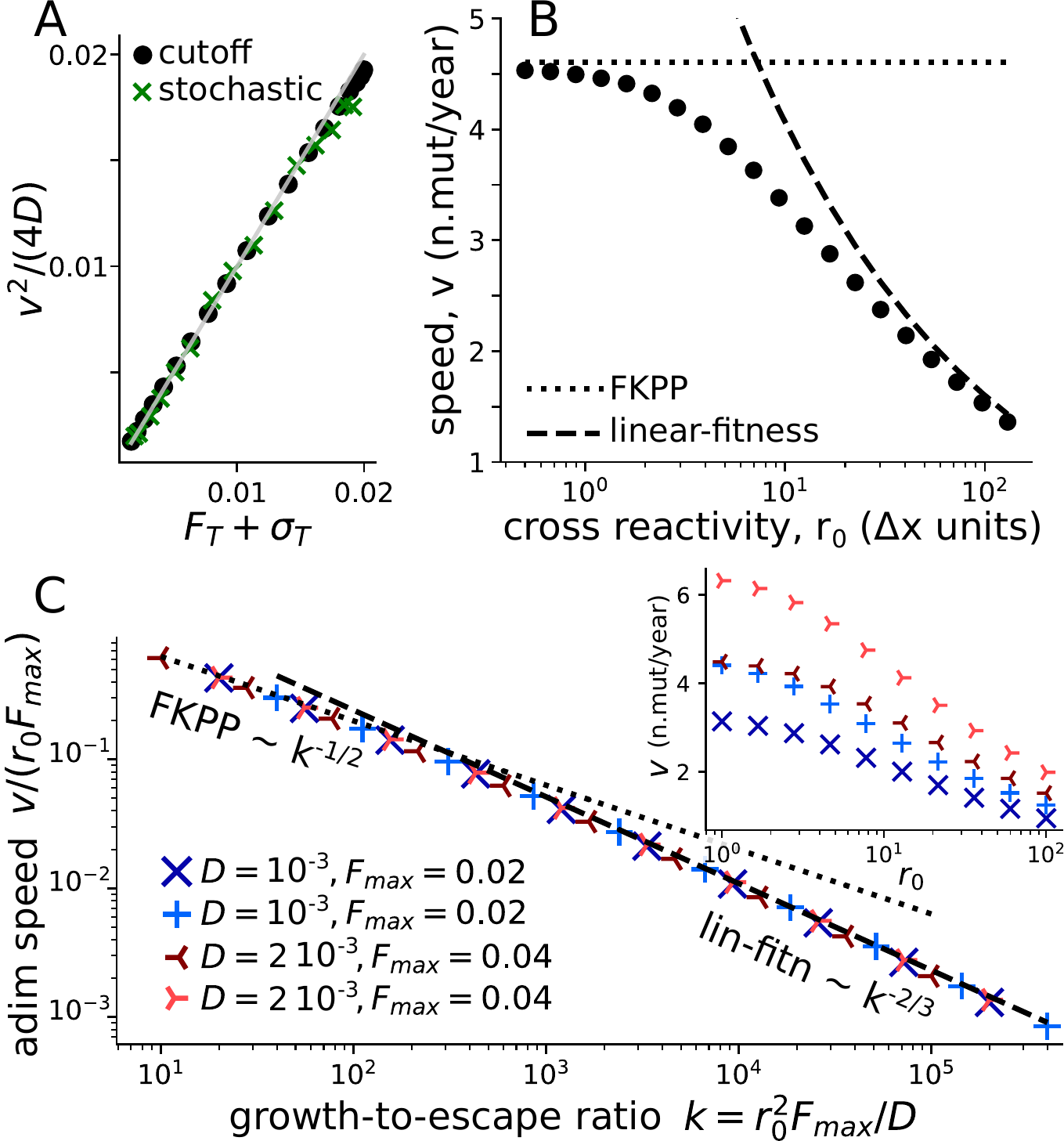}
	\caption{{\bf Wave speed.}
		(A): Numerical check of Eq.~\ref{eq:fitness_speed_rel} relating speed and fitness profile at the tip of the wave, for both the stochastic model and its deterministic approximation with a cutoff (Eq.~\ref{eq:dyn_n_cutoff}). Grey line is identity.
		Each point is defined by the values of speed, fitness and selection at the tip of the numerical solutions of our model.
		Different points are different values of $r_0$ and fixed values of the other parameters: \rev{$\mu_x=4 \; 10^{-3}$ $\text{day}^{-1}$, $\beta=0.12$ $\text{day}^{-1}$, $\gamma+\alpha=0.1$ $\text{day}^{-1}$, $M=5$, $N_h=10^{10}$}
		(B): Wave speed as a function of the cross reactivity, showing the crossover between the FKPP and linear-fitness regime.
		In the two extreme regimes, analytical predictions can be explicitly obtained (Eq.~\ref{eq:FKPP_speed} for the dotted line and Eq.~S24 for the dashed one).
		Same simulation parameters as (A).
		(C): Inset: speed {\em vs} cross-reactivity for different values of $D$ \rev{in n.mutations${}^2$ day${}^{-1}$} and \rev{$F_{\mathrm{max}} = \beta - \alpha - \gamma$} \rev{in day${}^{-1}$}. 
		The main plot shows the collapse of these curves as a function of the \rev{ growth-to-escape ratio} $k=r_0^2F_{\rm max}D$. The dashed and dotted lines are the theoretical predictions for the FKPP and linear fitness regimes.}
	\label{fig2}
      \end{figure}

We assume that Eq.~\ref{eq:dyn_n_cutoff} admits a stationary solution in the frame moving with constant speed $v$, so that all quantities depend on the reduced variable $u = x - vt$:
\beq\label{eq:u}
D \partial^2_u n(u) + v \partial_u n(u) + F(u)\Theta(n - n_c)n(u) = 0 .
\eeq
The dynamics of the wave is driven by its behaviour around the wave tip $u_T$, defined by $n(u_T)=n_c$, where we assume the fitness is locally linear, $F(u) \approx F_T + s_T (u - u_T)$.
This is a strong approximation which neglects everything that happens away from the tip, but, as shown below, it works extremely well, confirming the idea that the individuals at maximal fitness are the main drivers of evolution \cite{Desai2007}.
In that regime, Eq.~\ref{eq:u} may be solved exactly in the vicinity of $u_T$. The continuity condition between the $u>u_T$ and $u<u_T$ parts of the solution yields the relation ({\app} A.1):
\begin{equation}
\frac{v^2}{4 D} = F_T + \sigma_T,
\label{eq:fitness_speed_rel}
\end{equation}
where $ \sigma_T \equiv \xi_0 ( D s_T^2 )^{1/3}$, and $\xi_0 \approx -2.3381$ is the largest zero of the Airy function.
We verified numerically that this relation is satisfied for both the deterministic equation with a cut-off, Eq.~\ref{eq:dyn_n_cutoff}, and the original stochastic equation, Eq.~\ref{eq:dyn_n} (Fig.~\ref{fig2}A).
This relation connects the wave speed with the value of the fitness, $F_T$, and its derivative, $s_T$,  at the tip in a very general way, without assuming any specific behavior of the immune system. It can be potentially applied to every system showing traveling waves and a locally linearizable fitness at the tip. However, it is only implicit, since the position of the fitness tip $u_T$ itself needs to be computed from the model parameters in order to evaluate $F_T$ and $\sigma_T$. A second implicit equation for $v$ and $u_T$ may be obtained by imposing the normalization condition $\int du\,n(u)=N$ on the solution to Eq.~\ref{eq:u}, where $N$ is the number of infected hosts.

A special case is given by the regime of \quotes{small} cross reactivity (Fig.~\ref{fig1}B, left) where the fitness at the tip $F_T=\beta-\alpha-\gamma=F_{\rm max}$ is maximal, and its derivative $s_T=0$. Then Eq.~\ref{eq:fitness_speed_rel} is sufficient to determine the speed, giving back the classical expression for FKPP waves:
\begin{equation}
v = 2 \sqrt{F_{\rm max} D},
\label{eq:FKPP_speed}
\end{equation}
Fig.~\ref{fig2}B shows how the wave speed of our model converges to this limit (dotted line) for decreasing $r_0$.

In the opposite limit of a linear fitness profile, $F(u)\approx F(0)+su$ (Fig.~\ref{fig1}B, right), the normalization condition can be expressed analytically \cite{Tsimring1996, Cohen2005, Neher2013}, leading to the following approximated formula for the speed (see {\app} A.2 for a full expansion):
\begin{equation}
v \approx 2 \left( 3 s D^2 \ln\left(\frac{N}{n_c} \frac{s^{1/3}}{D^{1/3}} \right)\right)^{1/3}.
\label{eq:linear_fitness_speed}
\end{equation}
The fitness profile $F(u)$ may be obtained by integrating Eq.~\ref{eq:dyn_h}.
The stationarity condition of zero mean fitness, $F(0)=0$, gives an additional relation between $N$ and $v$, $N/N_h=vM(R_0^{1/M}-1)/r_0$. The gradient then reads $s = (\alpha + \gamma) M (R_0^{1/M} - 1)/r_0$.
This creates a closed system of 2 \rev{implicit} equations that allows us to estimate $N$ and $v$. The speed obtained numerically converges to that solution in the large $r_0$ regime (Fig.~\ref{fig2}B, dashed line).

To better understand the dependency of the crossover on the model parameters, we introduce the natural dimensionless parameter $k = {r_0^2 F_{\rm max}}/{D}$. 
It is equal to the ratio of two timescales: the typical time $r_0^2/D$ it takes a single virus to escape immunity by antigenic drift, and the characteristic doubling time $\propto F_{\rm max}^{-1}$ in absence of immunity.
\rev{We call this quantity the ``growth-to-escape ratio.''}
As Fig.~\ref{fig2}C shows, the normalized speed $v/(r_0F_{\rm max})$ collapses as a function of $k$ for a wide range of parameter values. 
The crossover takes place around $k \approx 10^{3}$.
In particular, a larger diffusion coefficient helps the virus to be well ahead of the immune coverage, which corresponds to the FKPP regime.
By \rev{contrast}, a large cross reactivity increases the immune coverage and pushes the system towards the linear-fitness regime.

\subsection{The evolutionary stable strategy has a crossover between maximizing the speed and the reproductive ratio}
\label{sec:es}

From this section on, we tackle the main question of this manuscript: understanding the evolutionary stable strategies of the viral population under immune pressure.
The first step is to ask whether a mutant competing with a resident population can displace it.
Consider a mutant strain with slightly different parameters than the resident one.
The evolution of its number, $n'(x,t)$, is given by Eq.~\ref{eq:dyn_n_cutoff}.
In the early days of this mutant, the resident strain is at stationary state, and that the mutant is too rare to contribute to the immune receptor density $h(u)$.

We assume that the fate of the mutant is determined by its behaviour at the tip of the wave, where the fitness profile is approximately linear. Then, as we show in {\app} C.1, the mutant population evolves in the moving frame as
$n'(u,t) = e^{\rev{\rho} t} \phi(u)$, with growth rate:
\begin{equation}
\rev{\rho} = F'_T + \sigma_T' - \frac{v^2}{4 D'},
\label{eq:invasion}
\end{equation}
where $F'_T=\beta'S(u_T)-\alpha'-\gamma'$ and $\sigma'_T=\xi_0(D's'_T)^{1/3}$, with $s'_T=\beta'\partial_uS(u_T)$, and
where the mutant parameters are indicated with a prime and $v$ is the speed of the resident population. The mutant invades if and only if $\rev{\rho}>0$.

As expected, when the mutant is phenotypically identical to the resident, $F_T'=F_T$, $\sigma'_T=\sigma_T$, $D'=D$, then Eq.~\ref{eq:fitness_speed_rel} implies $\rev{\rho}=0$, meaning that the mutant has no advantage or disadvantage.
We tested the validity of the invasion condition $\rev{\rho}>0$ for mutants of $\beta$ and $D$ in \SFig{S2}.
Note that this stability relation depends only on the fitness and the selection coefficient at the tip, without specific details of our immune framework.
This implies that it can be extended to other models.

Eq.~\ref{eq:invasion} allows us to see how the best viral strategy radically depends on the considered regime.
In the FKPP limit (small $r_0$), the condition becomes $F'_{\max}-v^2/4D'>0$, or equivalently $v'>v$ with $v'=2\sqrt{F_{\max}'D'}$:
The best strategy is to maximize the speed of adaptation.
In the linear-fitness regime (large $r_0$), the fitness in the bulk of the wave dominates Eq.~\ref{eq:invasion}, yielding the condition $F'(0)>0$ or equivalently $R_0'>R_0$: The best strategy is to maximize the reproductive ratio.

We can use Eq.~\ref{eq:invasion} to derive evolutionary stable points of the population in all regimes.
Consider phenotypic continuous variables $\theta$ over which the evolutionary process acts, so that $D(\theta)$, $\beta(\theta)$ and so on. The growth rate of an invading mutant, Eq.~\ref{eq:invasion}, depends on both the phenotypes of the resident and invading population, $\rev{\rho}(\theta';\theta)$. 
A stable point $\theta^*$ must satisfy $\rev{\rho}(\theta^*+\delta\theta;\theta^*)\leq 0$ for all perturbations $\delta\theta$. Since $\rev{\rho}(\theta^*;\theta^*)=0$, this implies $\partial_{\theta'}\rev{\rho}(\theta^*;\theta^*)=0$, which can be rewritten as ({\app} C.1):
\begin{equation}
\left.\partial_{\theta'} [(F_T' + \sigma_T')D']\right|_{\theta^*} = 0.
\label{eq:ES}
\end{equation}
We will  now use this condition to study the evolutionary stability of two distinct quantities independently: the mutation rate (next two sections) and the virulence (last section).

\subsection{Evolutionary stability of mutation rate under mutational load trade-off}

\begin{figure}
\includegraphics[width=.48\textwidth]{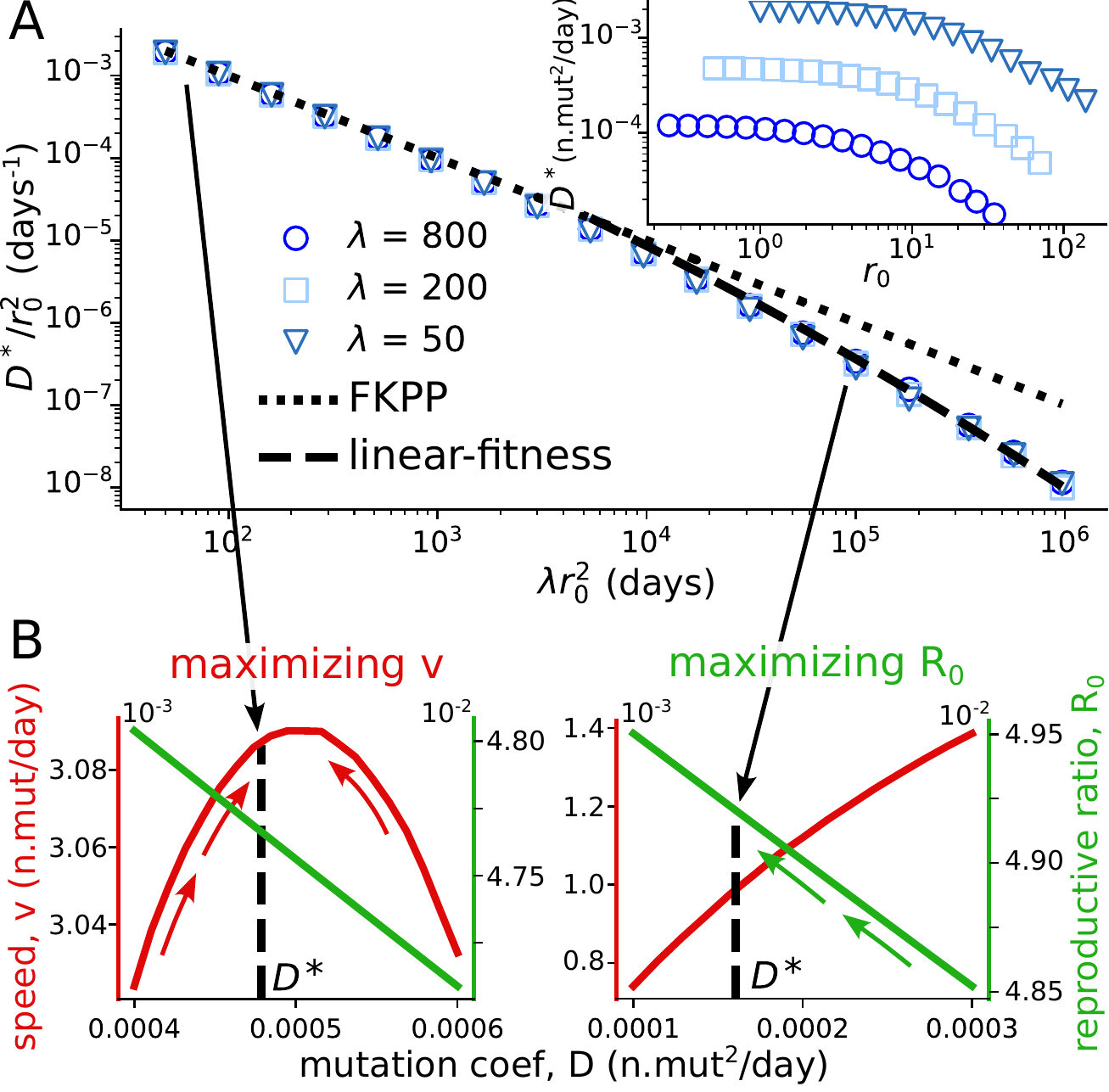}
\caption{{\bf Evolutionary stability of the mutation rate.} 
	(A): Evolutionary stable mutation coefficient $D^*$, re-scaled by $r_0^2$, as a function of the rescaled coefficient $\lambda r_0^2$.
	The dotted and dashed lines show the FKPP and linear-fitness predictions valid for small and large $r_0$ (\app 3B).
	Used parameters: \rev{$\beta_0=0.05~\text{day}^{-1}$, $\gamma+\alpha=0.04~ \text{day}^{-1}$, $M=5$, $N_h=10^{10}$. The value of $\lambda$ shown in the legend is in days/n.mutations${}^{2}$.
	Inset: same simulations plotted without rescaling. }
	(B): Wave speed and reproductive ratio as a function of the mutation coefficient for two extreme cross reactivities: \rev{$r_0=0.5$, $r_0=22.4$ at ${\lambda} = 200~ \text{ day/n.mutations}^2$.} The evolutionary stable coefficient is indicated with the black line. It tends to maximize the speed for small $r_0$, and the reproductive ratio for large $r_0$.}
\label{fig3}
\end{figure}

The mutation rate plays a key role for antigenic escape. By raising it, the ability of the virus to escape immune protection  increases, as shown by the positive dependency of the wave speed on $D$ (Eqs.~\ref{eq:FKPP_speed} and \ref{eq:linear_fitness_speed}).
However, a majority of mutations occurring in viruses are not affecting antigenic traits and generically decrease the intrinsic fitness of the strain \cite{Gabriel1993, Bull2007, Silander2007}. The larger the total mutation rate, the more these deleterious mutations accumulate and decrease the pathogen's infectivity, possibly leading to viral extinction.
In fact, increasing mutation rates is a widely used antiviral strategy \cite{Swanstrom2022}.
To account for this trade-off between the harmful effect of mutations and the benefits of antigenic escape, we let the infectivity depend on the rate of deleterious mutations per transmission $U_d$ as $\beta=\beta_0(1-U_d)$.
In {\app} B we derive this relation from the balance between mutation and selection \cite{kimura1966mutational, haigh1978accumulation, Bull2007} in an epidemiological context, using an approach similar to \cite{koelle2015effects}. The two rates $\mu_x$ and $U_d$ are assumed to scale both with the global mutation rate, so that they are linearly related, $U_d=a\mu_x$. This implies $U_d=\lambda D$, with $\lambda=2a/\Delta x^2$.

Using Eq.~\ref{eq:ES} with $D$ as the only phenotypic control parameter $\theta$, yields a implicit expression for the evolutionary stable state:
\begin{equation}
F_T^* (1 - 2 {\lambda} D^*) + \sigma_T^* \left( \frac{4}{3} - 2{\lambda} D^*  \right) - \gamma {\lambda} D^* = 0.
\label{eq:ES_mut_rate}
\end{equation}
In the two extreme limits $r_0\to 0$ and $r_0\to \infty$, we obtain explicit expressions that give us two different scalings between the normalized diffusion coefficient, and the normalized scaling factor: $D^*/r_0^2\sim (\lambda r_0^2)^{-1}$ for FKPP, and $D^*/r_0^2\sim (\lambda r_0^2)^{-3/2}$ for linear fitness ({\app} C.2). Note that $D^*/r_0^2$ may be interpreted as the inverse of the time it takes for a single strain to escape an immune protection by antigenic diffusion, while $\lambda r_0^2$ may be interpreted as the number of deleterious mutations accrued during that time.

These expressions are compared to numerical simulations of the evolutionary stability in Fig.~\ref{fig3}A, and confirm the scaling relation $D^*/r_0^2=f(\lambda r_0^2)$. We also tested the validity of the general stability condition, Eq.~\ref{eq:ES_mut_rate}, for both the stochastic model and its deterministic approximation with a cutoff \rev{(\SFig{S3})}.

Fig.~\ref{fig3}B shows the different behavior of the viral strategy depending on the value of $r_0$ discussed in the previous section: it tends to be the one that maximize the speed for small $r_0$, and the reproductive ratio for large $r_0$.

\subsection{Application to H3N2 evolution}
\label{sec:flu}

\begin{figure}
\includegraphics[width=0.48\textwidth]{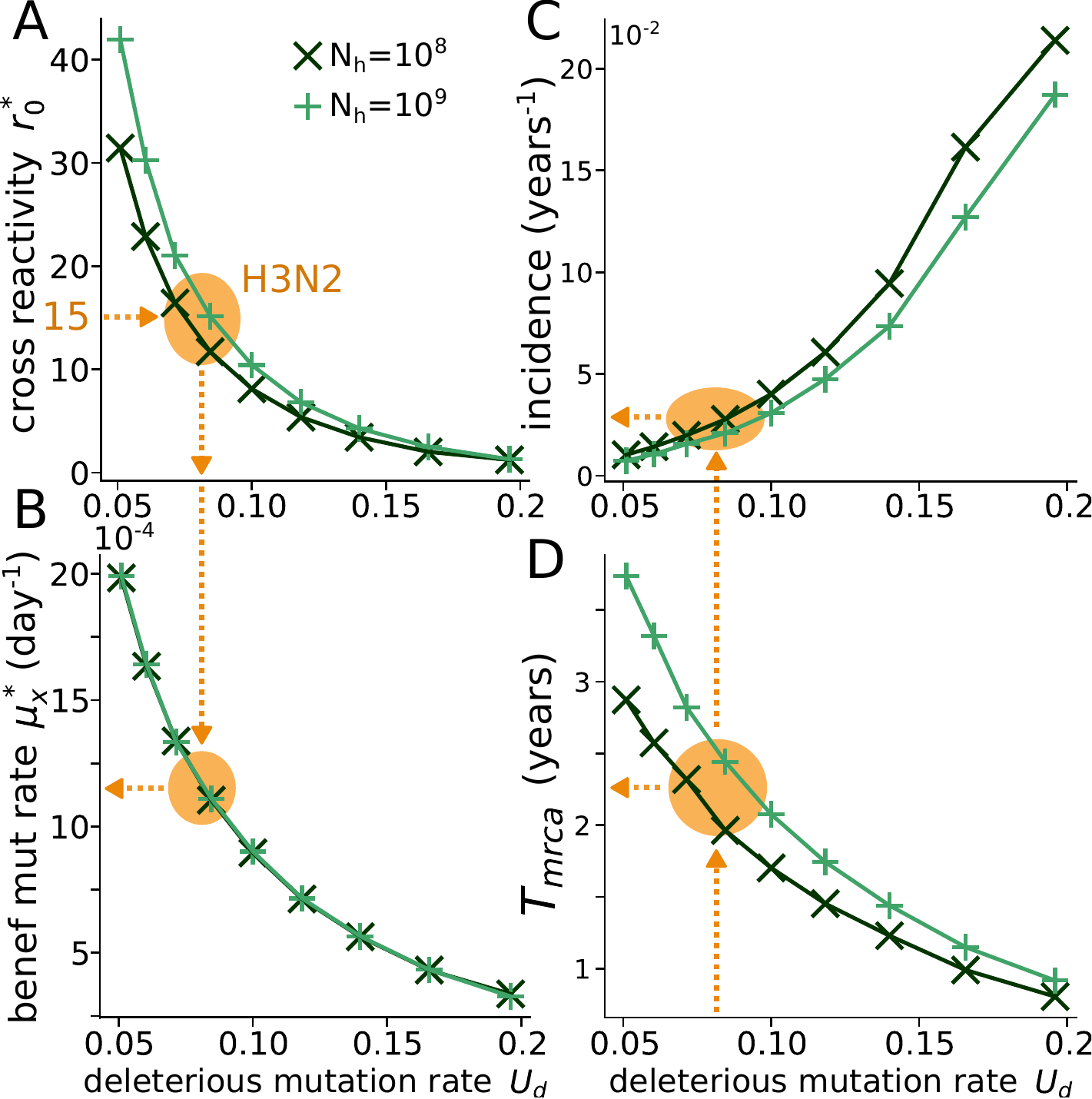}
\caption{{\bf Application to the H3N2 influenza strain}. Evolutionary
  stable values of the (A) cross-reactivity $r_0$ and (B) antigenic mutation \rev{rate $\mu_x$} as a function of the rate of  deleterious mutations $U_d$, for fixed
  values of the parameters $R_0 \approx 1.8$, $\gamma \approx 0.2$
  day$^{-1}$, $\alpha \approx 0$, $M=5$, and $N_h$ specified in the
  legend, with the additional constraint $v/\Delta x=2.6$ year$^{-1}$, as
  fixed by empirical estimation.
The orange arrows are the predictions of our model assuming using the
empirical estimate $r_0 \approx 15$ \rev{which implies $U_d \approx 0.08$.}
\rev{Panels (C) and (D) show the the incidence and the time from the most recent common ancestor.}
}
\label{fig4}
\end{figure}

We can apply the predictions from our model for the mutational trade-off to data obtained for the strain H3N2 of influenza infections.
Some of the parameters of the model can be fixed from data and their values are well-established in literature. Recall that we have set $\Delta x=1$.
The reproductive ratio is set to $R_0 = \beta_0(1-U_d)/(\alpha + \gamma)  \approx 1.8$,
 the recovery rate to $\gamma \approx 0.2$ day${}^{-1}$, and the virulence to $\alpha \approx 0$, which is negligible compared with the recovery rate.
We consider those parameters as fixed and we explore our model by varying the remaining ones, whose estimate is more indirect and less precise.
The number of receptors per host, $M$, is both difficult to estimate and specific to the particular modeling choice. However, we observed that results depend very weakly on its choice in the range $M=1$--$10$.
The effective population of hosts can be difficult to estimate, and we considered two reasonable choices $N_h=10^8, 10^9$.

We are left with three parameters to fix: the cross-reactivity in units of mutational steps $r_0$, the deleterious mutation rate $U_d$, and the diffusion coefficient $D$ \rev{(or equivalently $\mu_x = 2 D / \Delta x^2$)}.
The substitution rate of non-synonymous mutations in antigenically interacting regions of the virus, which can be identified with the wave speed in units of antigenic effects $v/\Delta x \approx 2.6$ year${}^{-1}$ \cite{biggerstaff2014estimates, Rouzine2018, einav2020snapshot}, imposes an implicit relation between these parameters.
In addition, the condition of evolutionary stability, Eq.~\ref{eq:ES_mut_rate}, imposes another constraint.
This leaves us with one degree of freedom, which we chose to control through the deleterious mutation rate, $U_d$. 

Figs.~\ref{fig4}\rev{A,B} shows the values of the two parameters $r_0^*$ and \rev{$\mu_x^*$} obtained by imposing the two conditions discussed above for different given values of $U_d$ (see \MM{sec:ES_and_speed} for details on how $r_0^*$ and \rev{$\mu_x^*$} are evaluated).
Hemagglutination inhibition data \cite{Smith2004} are consistent with a cross reactivity around $r_0 = 14-15$, as already used in previous work \cite{Luksza2014, Rouzine2018}.
From the left panel of Fig.~\ref{fig4} we can estimate $U_d \approx 0.08$ deleterious mutation per genome per transmission event, which is consistent with an independent estimate of $\approx 0.1$ \cite{koelle2015effects}.
Using the right panel of Fig.~\ref{fig4}, this value of $U_d$ in turn leads to an estimate for the \rev{beneficial mutation rate} $\mu_x\approx 1.2\cdot 10^{-3}$ antigenic mutations per day.

These numbers allow us to determine the regime of evolution of the virus. We estimate $k \approx 6 \cdot 10^4 \gg 10^3$, suggesting that H3N2 evolves in the linear fitness regime as has been assumed in other work, e.g. \cite{Marchi2021}.

\rev{
	We tested if other observables predicted by the model were consistent with empirical estimates.
	Fig.~\ref{fig4}C shows the incidence, defined as the fraction of infected people in a given time: $N\gamma / N_h$.
	The H3N2 strain infects $7-9\%$ of the population each year~\cite{Rouzine2018, wen2022potential}, while our model predicts $3-4\%$.
	Fig.~\ref{fig4}D shows the average time from the most recent common ancestor, which can be computed as the time the wave takes to reach its tip, $T_{MRCA} \approx c~u_T / v$, where $c \approx 1.66$~\cite{Neher2013, Marchi2021}.
	Our numerical estimates predict $2-2.5$ years, versus empirical estimates of $3.2 \pm 1.2$~\cite{Yan2019}.
For both quantities, our model captures the correct order of magnitude, recapitulating the overall features of influenza evolution with minimal ingredients.
}

\subsection{Evolutionary stability of virulence under transmission trade-off}

\begin{figure}
	\includegraphics[width=0.48\textwidth]{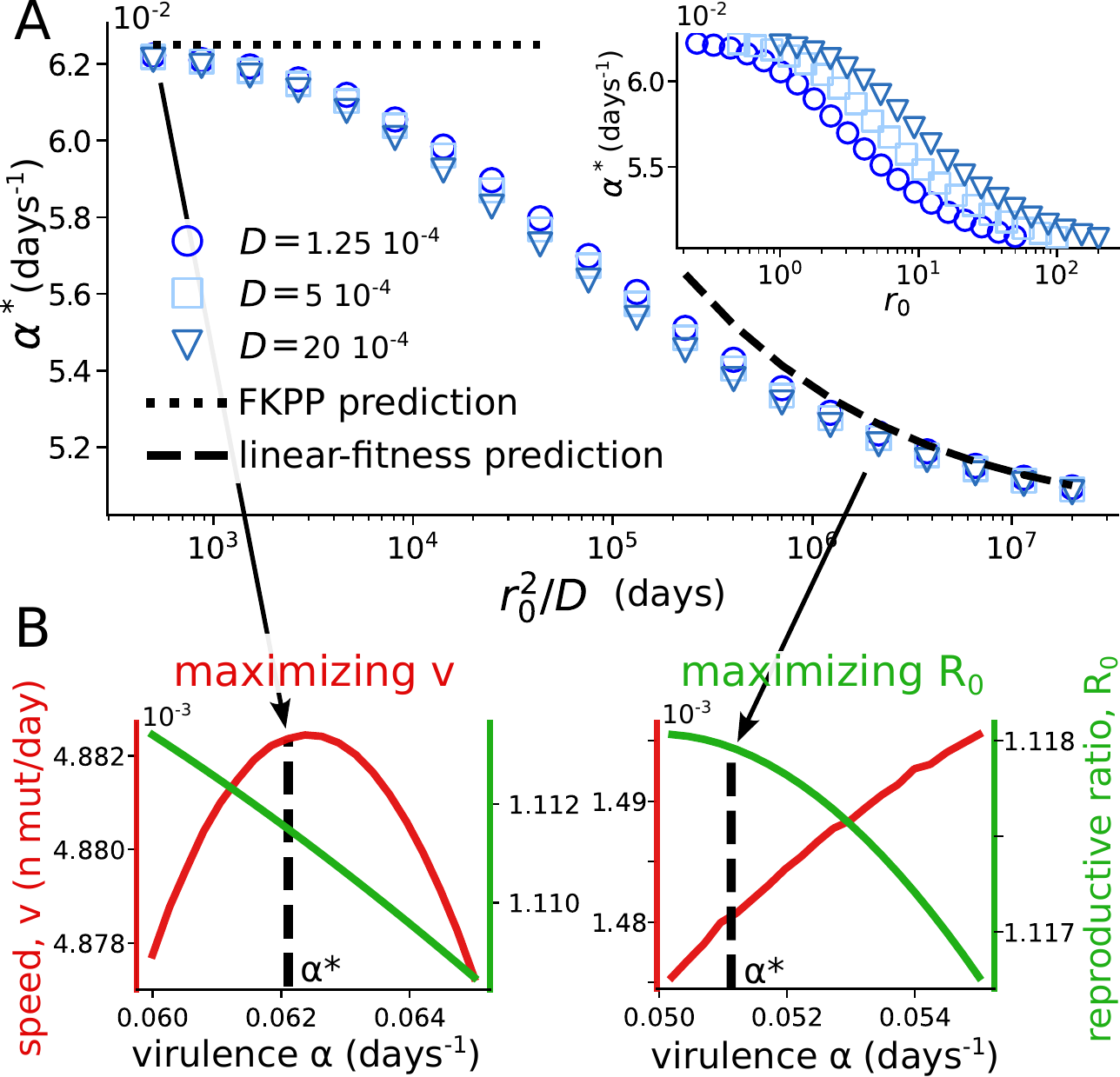}
	\caption{{\bf Evolutionary stability of virulence} (A):
    Evolutionary stable virulence as a function of the diffusion-escape time $D/r_0^2$. 
    Transmissibility has the form $\beta(\alpha) = b \sqrt{\alpha}$. The dotted and dashed lines show the two limits of FKPP and linear fitness (\app~3C).
	\rev{Parameters: $b=0.5~\text{days}^{-1/2}$, $\gamma=0.05~\text{days}^{-1}$, $M=5$, $N_h=10^{10}$.
	The values of $D$ in the legend is in n.mutations${}^2$/days.
	Inset: same simulations shown as a function of the non-rescaled variables.}
	(B): Wave speed and reproductive ratio as a function of the
        virulence for two different cross reactivities, \rev{ $r_0=0.5$ and $r_0=75.7$, and $D=5~10^{-4}~\text{n.mutations}^2\text{/days}$.} }
	\label{fig5}
\end{figure}

We now turn to the application of our stability condition to the evolution of the virulence $\alpha$. 
Recall that according to the classical argument (which ignores immune escape and waves), virulence should evolve to maximize the viral reproductive ratio $R_0=\beta/(\gamma+\alpha)$ \cite{May1990, Alizon2009}. If $\beta$ is an increasing but concave function of $\alpha$, there exists a tradeoff $\alpha^*$ between the opposing needs to increase transmissibility and to decrease virulence.

By contrast, applying Eq.~\ref{eq:ES} with $\theta=\alpha$ gives us the following condition for the evolutionary stable virulence:
\begin{equation}
  \left(F_T+\alpha+\gamma+\frac{2}{3}\sigma_T\right)\beta^{-1}\partial_\alpha\beta =1.
  \label{eq:ES_virulence}
\end{equation}
To check the validity of this relation, we numerically looked for the evolutionarily stable value of the virulence $\alpha^*$ as a function of $r_0$, \rev{for the commonly used concave function} $\beta(\alpha) = b \sqrt{\alpha}$ (Fig.~\ref{fig5}A). The numerical results are consistent with Eq.~\ref{eq:ES_virulence} (\rev{\SFig{S4}}), and show a collapse of $\alpha^*$ as a function of $r_0^2/D$.

As we expect from previous arguments, the evolutionary endpoint maximizes the speed of the wave $v$ in the FKPP regime of low cross-reactivity  ($\partial_\alpha\beta=1$), consistent with a previous analysis \cite{Sasaki2021}. By contrast, the reproductive ratio $R_0$ is maximized in the linear-fitness regime of high cross-reactivity, consistent with the classical result in absence of escape.
The two limit cases are illustrated in Fig.~\ref{fig5}B.

Fig.~\ref{fig5}A also implies that short cross-reactivities favor the evolution of higher virulence. 
\rev{
This result holds for every concave function $\beta(\alpha)$.}
This highlights the importance of correctly estimating the quantitative effect of immunity for predicting the evolution of non-antigenic traits.

\section{Discussion}
The relevance of propagating waves for describing the continual dynamics of viral escape from immunity according to the ``Red Queen'' hypothesis has long been recognized, but using two seemingly contradictory mathematical paradigms. The FKPP wave, originally introduced to describe the spatial spreading of beneficial mutations \cite{Fisher1937}, was shown to emerge in simple models of joint viral-immune dynamics along an antigenic dimension \cite{Sasaki1994}. The fitness-wave model was proposed to model the evolution of RNA viruses with an infinite reservoir of beneficial mutations \cite{Tsimring1996}. Shifting immunity was then proposed as a mechanism for such a reservoir in models of viral-immune co-evolution \cite{Marchi2021,Yan2019}. The two descriptions lead to markedly different predictions for the rate of adaptation and its dependence upon the antigenic mutation rate and host population size. We have explicitly showed that these two descriptions correspond to limiting cases of the same model, reconciling apparently \rev{incompatible} approaches.

These two limits may be understood intuitively as follows. In the regime of low cross-reactivity, where the FKPP wave description holds, the fittest variants at the tip of the wave grow in an almost entirely susceptible host population. They all have comparable growth rates, and so do variants with additional escape mutations, which only confer a negligible advantage until hosts start getting infected in large numbers. By contrast, in the regime of high cross-reactivity, where the fitness-wave description is valid, even the most advanced escape mutants face a partially immune host population. This implies that additional escape mutations at the tip have an immediate growth advantage over their ancestors, driving the evolution of the virus.

Applying our theory to the evolution of H3N2 influenza, by plugging parameter estimates along with the \rev{additional} assumption that the viral mutation rate is evolutionary stable, we 
\rev{are able to recover the empirical estimates of the incidence rate and the time from the most recent common ancestor, which have not been used in the fitting procedure.
We} 
argued that H3N2 falls in the linear-fitness regime, where cross-reactivity is relatively large.
This implies that the new variants driving influenza evolution are still largely subject to the hosts' collective immunity, albeit a bit less so \rev{that they retain a small fitness advantage to the the majority of circulating strains}. This is consistent with the observation that emerging variants have a moderate effective reproductive number relative to the basic one $R_0$ \cite{Biggerstaff2014}. We may expect the evolution of SARS-CoV-2 to fall in the same regime as it settles in its endemic state.

{\VC
We showed that our model is able to capture key features of influenza evolution, and allows us to estimate both evolutionary and epidemiological quantities such as the antigenic mutation rate and the incidence rate. However, this model oversimplifies several aspects of influenza evolution, and notably ignores the rare mutations capable of inducing large jumps in the antigenic space~\cite{Smith2004, koel2013substitutions}. While our continuous model is a simple approximation to influenza evolution, we believe that a more detailed treatement of its mutational process~\cite{good2012distribution} could provide a more accurate estimation of the parameters presented in this paper, and would also allow for the estimation of yet unaccessible parameters such as the rate and the size of large antigenic jumps.
}

We showed that the cross-over between {\VC high and low cross-reactivity} has a strong impact of the evolution of non-\rev{antigenic} traits. In the low cross-reactivity regime, the ESS maximizes the speed of the wave (the rate of \rev{adaptation}), consistent with \cite{Sasaki2021}: strains that get ahead  at the tip of the wave outcompete slower ones.
In the high cross-reactivity regime however, the ESS maximizes the reproductive number. Intuitively, all strains are under strong immune pressure, so that their effective growth rate is close to 0, with a minute growth advantage for the most advanced immune-escape strains; any intrinsic fitness advantage (larger $R_0$) is likely to fix in the bulk of the wave, regardless of the wave's speed.
This last conclusion differs from that of Ref.~\cite{Sasaki2021}, where it was argued that the speed of the wave was maximized in the ESS regardless of the extent of cross-immunity. This discrepancy may be explained by the different assumptions about the dynamical regime. In this paper we specifically looked for strict steady-state solutions (in the moving frame of the wave), while Ref.~\cite{Sasaki2021} also considered oscillatory solutions, which emerge in the high cross-reactivity limit (where we claim the fitness-wave solution holds). One limitation of our approach is that it ignores the possible effect of such oscillations. However, oscillations also lead to near population collapse, and may not survive a full stochastic setting where extinction is likely. Which dynamical regime is relevant for real viruses remains an interesting question for future research.

Our results have several implications for the evolution of non-antigenic traits.
The suggestion that respiratory viruses may be in the linear-fitness regime implies that their mutation rate should evolve towards low values to minimize their mutational load, at the expense of their ability to escape immunity. More broadly, our result that $R_0$ is maximized in the linear-fitness regime implies that antigenic and non-antigenic evolutions are decoupled, suggesting that previous arguments that ignored antigenic escape may still be valid. We also predict that viruses for which there is more cross-immunity should evolve to be less virulent.

\section{Materials and methods}

\subsection{Deterministic simulation with a cut-off}
\label{sec:simulations}

Eqs.~\ref{eq:dyn_h}--\ref{eq:dyn_n_cutoff} are simulated using the Euler–Maruyama method. The code can be found at the following repository: \url{https://github.com/statbiophys/viral_coevo}.
Time and space are discretized with resolution $\delta t$ and $\delta x$ respectively \rev{(note that $\delta x$ is a \quotes{parameter} of the algorithm for solving the continuous equation and should not be confused with $\Delta x$)}. 
We choose $\delta x$ small compared to both $r_0$ and the width of the wave (the second condition between checked \textit{a posteriori}), and then set $\delta t$ to satisfy the Courant-Friedrichs-Lewy condition, $D \delta t / \delta x^2 < 1$.

The one-dimensional antigenic space is simulated with periodic boundary conditions, with box size larger than the immune persistence $v M N_h/N$. Previous passages are erased by setting $h(x,t)$ to zero ahead of the wave. 
\rev{In some regimes, the immune protection cannot be sufficient to prevent individuals at the back of the wave to grow again, creating a secondary wave. 
Since the behavior of the primary wave is not affected and secondary waves can create numerical instabilities we artificially impose perfect immunity ($S=0$) at the back of the wave.}

For large cross-reactivities $r_0$, some initial conditions may lead to oscillations occurring around the stationary state of the wave \cite{Sasaki2021}, leading to extinctions when a cut-off is imposed.
To start close to the stationary wave solution, we initialize $n(x,t)$ as a skewed Gaussian:
$n(x, 0) = k_n e^{-x^2/2} \left(1 + \text{erf}\left(4 x/\sqrt{2} \right) \right)$,
where the normalization coefficient $k_n$ is chosen to obtain an incidence rate of $1\%$: $\sum_x n(x,0) \; \delta x = N_h/100$.
The immune protection is initialized as:
$h(x, 0) = k_h H(x) e^{-|x| / \rho}$,
where $H(x)=1$ if $x>0$ and $0$ otherweise, $\rho = r_0(R_0^{1/M} - 1)^{-1}$, and $k_h$ is chosen such that $\sum_x h(x,0) \; \delta x = N_h M$.
To study the stationary state, we first simulate for some time $T_{\rm burn\, 1}=100$--$1000$ days without the cutoff ($n_c = 0$), and some time $T_{\rm burn\, 2}=200$--$20,000$ days with the cut-off (the larger the $r_0$, the longer the equilibration time).

\subsection{Stochastic simulation}
\label{sec:stoch_sim}

The stochastic simulation of Eqs.~\ref{eq:dyn_n}--\ref{eq:susceptibility} uses a hybrid deterministic-stochastic approach. The bulk of the wave, characterized by very large numbers where demographic noise is negligible, is treated deterministically, while the front is simulated stochastically.
We fix a threshold on the number of infected people $n_{\rm stoch}$ ($=10^6$ for Fig.~\ref{fig2}, $10^4$ for Fig.~\ref{fig3}, $10^5$ for Fig.~\ref{fig5}).
If $n(x,t) \rev{\delta x} < n_{\rm stoch}$, we update its value at each time step following the following stochastic prescription, which appropriately models demographic noise.
Each individual can mutate with probability $p = 1 - \exp[-2 D \delta t / \delta x^2]$, generating a binomial number of mutations $N_{\rm mut,r}(x) \sim \mathrm{Binom}(n(x)\delta x, p/2)$ to the right at $x+\delta x$, and a similarly distributed number $N_{\rm mut,r}(x)$ to the left at $x-\delta x$.
Numbers are then updated as: $n(x, t+\delta t/2) = n(x, t) + \delta x^{-1}[ N_{\rm mut,l}(x+\delta x) -N_{\rm mut,l} (x)+ N_{\rm mut,r}(x-\delta x) -N_{\rm mut,r}(x)]$.
Growth is then implemented by updating
$n(x, t+\delta t) \sim \mathrm{Poiss}(\bar{N}(x)) / \delta x$, with $\bar{N}(x) = (1  + F(x,t)\delta t) n(x,t+\delta t/2) \delta x$.

\subsection{Evolutionary stability simulations}
\label{sec:ES_simulations}

To simulate the evolution of populations with non-antigenic mutations, we consider a two dimensional system $(x,\theta)$, where $\theta$ is the phenotypic parameter over which evolution is acting (i.e. $D$ or $\alpha$ in the two examples considered in this paper).
We then assume that the system diffuses slowly with coefficient $\epsilon$ in the second dimension, and the full simulated equation reads:
\begin{equation*}
\begin{aligned}
\partial_t n(x,\theta,t) = \;F(x,\theta,t) n(x,\theta,t) + D(\theta) \partial^2_x n(x,\theta,t) \\
\epsilon \partial^2_\theta n(x,\theta,t) + \text{demographic noise}
\end{aligned}
\end{equation*}
where the demographic noise can be treated fully or through a cut-off as in Eq.~\ref{eq:dyn_n_cutoff}.
The diffusion coefficient $\epsilon$ is chosen to be as small as possible, so that the dynamics of the wave be much faster than the evolutionary time scale over which $\theta$ changes, and that the simulation converge to an evolutionary stable state well peaked in $\theta$ . The $\theta$ dimension is discretized with step $\delta \theta$.
After the simulation has converged, we take as the evolutionary end point $\theta^*=\langle \theta\rangle=\int dx d\theta\, n(\theta,x) \theta$.

\subsection{Imposing evolutionary stability and speed of the wave for the \rev{H3N2} study}
\label{sec:ES_and_speed}

Here we detail the method for getting $r_0^*$ and $D^*$, all other \rev{parameters} ($R_0$, $\gamma$, $M$, $N_h$, $U_d$) being fixed, by using the following two relations: $v/\Delta x=2.6$ years$^{-1}$, and evolutionary stability.

The pseudo-algorithm for finding these two values is the following:
\begin{itemize}
\item[1] Choose an initial guess for $r_0$.
	\item[2] Find $D^*$ that matches the speed condition through nested iteration:
          \begin{itemize}
          \item[a] Choose an initial guess $D$.
          \item[b] Calculate the speed $v$ with $F(x)=\gamma(R_0 S(x)-1)$.
          \item[c] Update $D^*\leftarrow D^*-\alpha_1(v-\hat{v})$, where $\hat{v}=2.6$ is the target value, and $\alpha_1$ is a learning rate. Go to b.
          \end{itemize}
          \item[3] Run an evolutionary simulation where $D$ is left free as in Sec.~\ref{sec:ES_simulations}, now with $F(x)=\gamma(R_0 S(x)(1- \lambda D)/(1-U_d)-1)$, with $\lambda=U_d/D^*$ fixed. Call $D'$ the resulting evolutionarily stable point.
          \item[4] Update $r_0\leftarrow r_0-\alpha_2(D'-D^*)$ and go back to 2.
          \end{itemize}

\section*{Acknowledgements}
The study was supported by the European Research Council COG 724208
and ANR-19-CE45-0018 ``RESP-REP'' from the Agence Nationale de la Recherche and DFG grant CRC 1310
``Predictability in Evolution''.

\bibliographystyle{pnas}

\appendix
\onecolumngrid

\newpage

\renewcommand{\thefigure}{S\arabic{figure}}
\setcounter{figure}{0}


 \onecolumngrid

\section{Derivation of the wave speed}

\subsection{Fitness speed relation}
\label{sec:fitness_speed_rel}

To derive the speed of the viral wave, we look for solutions of the form $n(x,t) = n(x - v t, t)$.
To this end, we consider Eq.~5 in the frame of reference of the wave, $u = x - v t$, and we assume that it admits a stationary solution
\begin{equation}
\label{SM_eq0}
D \partial^2_u n(u) + v \partial_u n(u) + F(u)\Theta(n - n_c)n(u) = 0 .
\end{equation}

The main assumption is that the behavior of the wave is driven only by the individuals at the front tip.
If we define $u_T = x_T - v t$ the antigenic position at which $n(u_T) = n_c$, we encode this assumption by considering only the behavior around $u_T$.
In particular, the fitness is approximated is a linear function around it: $F(u) \approx F(u_T) + \partial_u F(u_T) (u - u_T) \equiv F_T + s_T (u - u_T)$.

The next step is to solve equation~\ref{SM_eq0} on the right and on the left of $u_T$, and then to impose the continuity of the infected host profile and its derivative.
For $u > u_T$ the equation reduces to
\begin{equation}
D \partial^2_u n(u) + v \partial_u n(u) = 0 .
\end{equation}
solving this equation and imposing $n(\infty) = 0$ and $n(u_T) = n_c$, one can find
\begin{equation}
n(u) = n_c \exp \left( - \frac{v}{D} (u - u_T) \right)  \hspace{0.5cm} \text{for} \hspace{0.5cm} u > u_T.
\label{eq:SM_n_right}
\end{equation}

The behavior of the wave on the left of the tip can be obtained by solving
\begin{equation}
\label{SMeq1}
D \partial^2_u n(u) + v \partial_u n(u) + (F_T + s_T (u - u_T))n(u) = 0 .
\end{equation}
We consider the solution as $n(u) = \exp(-v u/(2D)) \psi(u)$.
By plugging it into Eq.~\ref{SMeq1}, the equation to solve becomes
\begin{equation}
\frac{D}{s_T} \partial^2_u \psi(u) + \left(\frac{F_T}{s_T} - u_T - \frac{v^2}{4 D s_T} + u \right) \psi(u) \equiv c_1 \partial^2_u \psi(u) + \left(u - c_2\right) \psi(u) = 0 ,
\label{eq:SM_Airy}
\end{equation}
where the coefficients $c_1 = D/s_T$ and $c_2 = u_T + v^2/(4Ds_T) - F_T/s_T$ are introduced for a shorter notation.
One can then realize that the equation can be rewritten as an Airy equation with a simple change of variable $y = c_2 - u$.
The solution can be then expressed as a linear combination of Airy functions:
\begin{equation}
\psi(y) = A\text{Ai}\left(y c_1^{-1/3}\right) + B\text{Bi}\left(y c_1^{-1/3}\right) .
\end{equation}
However, by knowing that the function decays to zero for $u \rightarrow \infty$, the coefficient $B$ can be set to zero.
This leads to the following solution:
\begin{equation}
n(u) = A \exp \left( - \frac{v}{2D} u  \right) \text{Ai}\left((c_2 - u) c_1^{-1/3}\right)  \hspace{0.5cm} \text{for} \hspace{0.5cm} u < u_T.
\label{eq:SM_n_left}
\end{equation}

\begin{figure*}
	\includegraphics[width=1\textwidth]{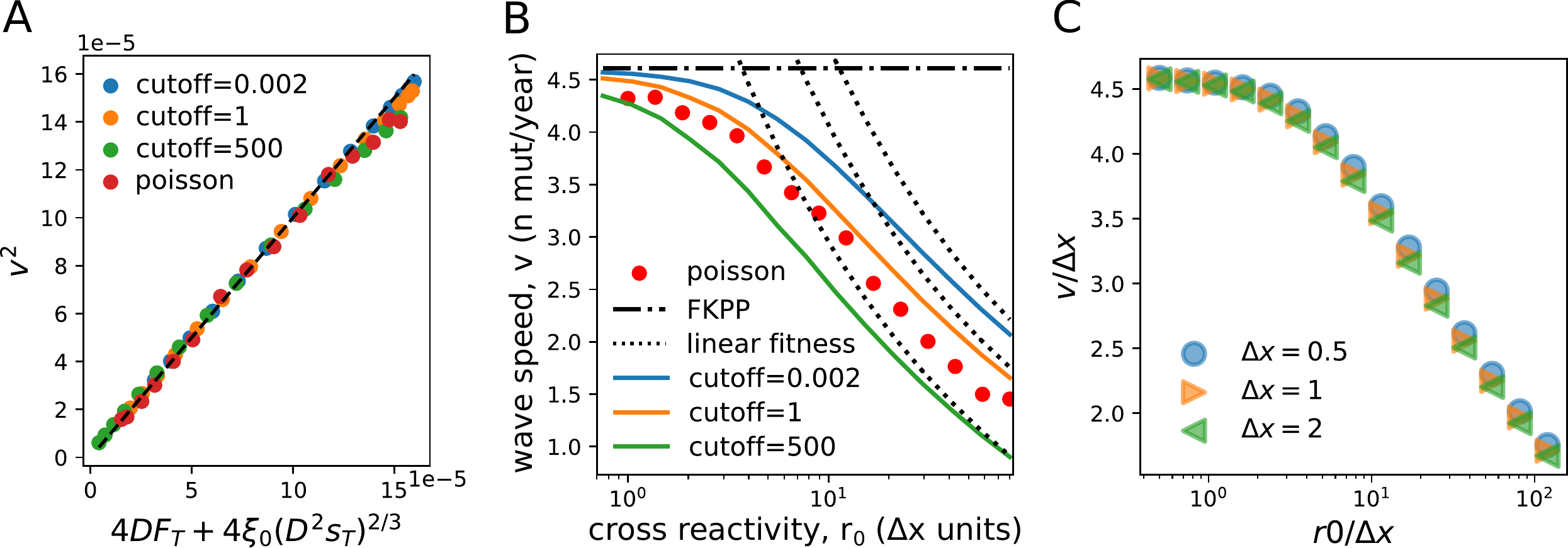}
	\caption{(A): Testing the fitness speed relation, Eq.~\ref{eq:SM_fitness_speed_rel} for different deterministic simulations of Eq.~5 having different cutoffs $n_c$.
	The relation is tested also for a stochastic simulation (red dots) whose details are described in Sec.~3B.
	The threshold below which the simulation is fully stochastic is $n_{stoch} = 10^5$.
	In this latter setting, speed, fitness and selection at the tip fluctuates, and to build the figure are averaged over a stationary trajectory.
	The other parameters of the simulations are: \rev{$\mu_x=4 \; 10^{-3}$ $\text{day}^{-1}$, $\beta=0.12$ $\text{day}^{-1}$, $\gamma+\alpha=0.1$ $\text{day}^{-1}$, $M=5$, $N_h=10^{10}$}.
	The panel (B) shows the wave speed as a function of $r_0$.
	The stochastic setting behaves like a deterministic one with a proper cutoff, in this case around $n_c \sim 10$.
	Panel (C) tests that the value of the speed is approximately invariant for different values of $\Delta x$ if the antigenic space is taken in units of $\Delta x$.
	This confirms that Eq.~5 is approximately invariant by change of spatial units of measures, and allows us to fix them by choosing $\Delta x = 1$. 
	}
	\label{fig:SM_speed_stoch}
\end{figure*}

Now we have to impose continuity and equality of the first derivative between the Eq.~\ref{eq:SM_n_right} and Eq.~\ref{eq:SM_n_left} at the intersection point $u_T$.
After some algebra, the two conditions lead to the following expression:
{\VC 
\begin{equation}
\frac{\text{Ai}'\left((c_2 - u_T) c_1^{-1/3}\right)}{\text{Ai}\left((c_2 - u_T) c_1^{-1/3}\right)} = \frac{v }{2 D} c_1^{1/3} = \frac{v}{2(D^2s_T)^{1/3}}.
\end{equation}
This dimensionless ratio diverges in the FKPP regime as the slope of the fitness profile vanishes, but could reach more moderate values in the linear fitness regime. In a first approximation we assume that it is large enough for the  Airy function at the denominator on the left-hand-side to be close to its first zero $\xi_0 \approx -2.3381$. We can then expand the function around $\xi_0$:
\begin{equation}
\frac{\text{Ai}'\left( \xi_0 + \epsilon \right)}{\text{Ai}\left( \xi_0 + \epsilon \right)} \approx \frac{\text{Ai}'\left( \xi_0 + \epsilon \right)}{\text{Ai}\left(\xi_0\right) + \epsilon \text{Ai}'\left( \xi_0 \right) } \approx \frac{1}{\epsilon} = \frac{v }{2 D} c_1^{1/3} .
\end{equation}
The value of $\epsilon$ can be found as $\epsilon + \xi_0= (c_2 - u_T) c_1^{-1/3}$ and, after some algebra, one can obtain
\begin{equation}
c_2 - u_T - \xi_0 c_1^{1/3} = \frac{2D}{v}.
\end{equation}
Finally, by expliciting the coefficients $c_1$ and $c_2$, a relation between the speed of the wave, the fitness at the tip and its derivative can be obtained:
\begin{equation}
\frac{v^2}{4 D} = F_T + \xi_0 \left( D s_T^2 \right)^{1/3} + \frac{2D s_T}{v}.
\label{eq:SM_fitness_speed_rel_gen}
\end{equation}
The third term on the right handside isn't a priori negligible in the linear-fitness regime for which we show in the next section that $v \sim (D^2 s_T)^{1/3}$. However, we tested this fitness speed relation for different values of the cutoff $n_c$  in Fig.~\ref{fig:SM_speed_stoch}A and observed that neglecting this term provides a very good approximate relation:
\begin{equation}
\frac{v^2}{4 D} \simeq F_T + \xi_0 \left( D s_T^2 \right)^{1/3}.
\label{eq:SM_fitness_speed_rel}
\end{equation}
A more thorough justification requires to calculate the exact speed in the linear-fitness regime to estimate the relative contribution of each term and is provided in the next section. Overall, Fig.~\ref{fig:SM_speed_stoch}A shows that Eq.~\ref{eq:SM_fitness_speed_rel} is satisfied independently of the cutoff as well as for the stochastic model described in Sec.~4B.}

We want to stress that the expression Eq.~\ref{eq:SM_fitness_speed_rel} still depends on one unknown quantity: the position of the tip, $u_T$. As discussed also in the main text, this makes the expression only implicit and does not provide a direct prediction of the wave speed from the model parameters. However, since we do not specify the shape of $F(u)$, the validity of the equation goes beyond the presented model and can be a valuable result for other frameworks that study traveling wave dynamics, connecting the wave speed with interpretable quantities, i.e. $F_T$ and $s_T$.

In general, to close this expression, one first step is to impose the normalization of the population profile
\begin{equation}
	\int_{-\infty}^{\infty} du \; n(u) = N .
	\label{eq:SM_N}
\end{equation}
This integral needs the value of the function $n(u)$ in the whole domain, which, in our case, is unknown and does not allow us to close the expression for the wave speed.
However, this can be done in the extreme regimes (see also the main text): in the FKPP regime Eq.~\ref{eq:SM_fitness_speed_rel} looses its dependence on $u_T$ and does not require Eq.~\ref{eq:SM_N}, while, in the linear-fitness regime, the integral of Eq.~\ref{eq:SM_N} can be solved.
In this latter case $N$ is still unknown, but the fitness depends on it and the chain of conditions can be closed by imposing that the average fitness is zero (see next section for the derivation).

However, in general, the value of the speed does depend on the cutoff value, as shown by Fig.~\ref{fig:SM_speed_stoch}B where the speed is plotted as a function of the cross reactivity.
This dependency is weak, as proven by the fact that varying the cutoff for several order of magnitude leads to quite similar speeds
In the two limit cases, this dependency can be analytically understood.
In the FKPP regime, there is no dependency on $n_c$, while, in the linear-fitness case, the cutoff appears as a factor dividing the population size, Eq.~\ref{eq:SM_linear_fitness_speed}.
The stochastic simulations show a speed which is compatible with an effective value of the deterministic cutoff.

\subsection{Wave speed in the linear-fitness regime}
\label{sec:linear_fitness}

Here we consider the system in the linear-fitness regime, where the wave feels an approximately linear fitness profile.
This regime is obtained for large values of $r_0$ or, more precisely, for a small adimensional coefficient $k \ll 10^{-3}$.
In such a condition we assume that the fitness is linear and zero at
the center of the wave: $F(u) = s u$ \rev{(so that $u=0$ is the mean
  viral position in the co-moving frame)}.
The explicit expression of the fitness slope $s$ will be found later.
This allows us to write down the approximation of Eq.~5, which we consider at stationarity in the frame of reference of the moving wave
\begin{equation}
	D \partial^2_u n(u) + v \partial_u n(u) + s u \Theta(n - n_c)n(u) = 0 .
	\label{eq:SM_dyn_lin_fitn}
\end{equation}
As for the previous derivation of the fitness-speed relation, we want to solve the equations on the right and on the left of the tip of the wave $u_T$, where $n(u_T) = n_c$, and then impose the continuity of the function and its derivative on the junction point.
On the right side, $u > u_T$ the solution is exactly equal to Eq.~\ref{eq:SM_n_right}.
For $u < u_T$, the structure of the equation is the same as the previous section, but with different coefficients.
The solution is then the Airy function~\ref{eq:SM_n_left} with $c_1 = D/s$ and $c_2 = v^2/(4 D s)$:
\begin{equation}
n(u) = A \exp \left( - \frac{v}{2D} u  \right) \text{Ai}\left((c_2 - u) c_1^{-1/3}\right)  \hspace{0.5cm} \text{for} \hspace{0.5cm} u < u_T.
\label{eq:SM_lin_fitn_n}
\end{equation}
Importantly, this population profile is valid in the whole antigenic space, while before we were considering only the expression close to the tip.
As before, we can impose the the continuity of the function and its derivative in $u_T$ and get
\begin{equation}
	c_2 - u_T - \xi_0 c_1^{1/3} = \frac{v^2}{4 D s} - u_T - \xi_0 \left( \frac{D}{s} \right)^{1/3} = \frac{2 D}{v}.
	\label{eq:SM_uT}
\end{equation}

This provides a first equation connecting the speed with the model parameters, however we still have the unknown $u_T$.
To find a second condition for fixing its value, we consider the normalization of the host population
\begin{equation}
	\int_{-\infty}^{u_T} du \; n(u) \approx N,
\end{equation}
where we consider the contribution to $N$ given by the right side of the cutoff negligible.
The expression of the number of hosts for $u<u_T$ is known in this regime, Eq.~\ref{eq:SM_lin_fitn_n} (with the previously specified coefficients $c_1$ and $c_2$).
This leads to the following integral
\begin{equation}
	\frac{n_c}{\text{Ai}\left((c_2 - u_T) c_1^{-1/3}\right)} \int_{-\infty}^{u_T} du \; \exp \left( - \frac{v}{2D} (u - u_T)  \right) \text{Ai}\left((c_2 - u) c_1^{-1/3}\right) \approx N,
\end{equation}
where the coefficient $A$ in Eq.~\ref{eq:SM_lin_fitn_n} has been fixed with $n(u_T) = n_c$.

By making a change of variable in the integral $\xi = \xi_0 + (u_T - u)/c_1^{1/3}$ and using expression \ref{eq:SM_uT} one obtains
\begin{equation}
	\int^{\infty}_{\xi_0} d\xi \; c_1^{1/3} \; \exp \left( \frac{v c_1^{1/3}}{2D} \xi  \right) \text{Ai} \left(\xi + \frac{2D}{v c_1^{1/3}}\right) =
	\frac{N}{n_c} \exp \left( \frac{v c_1^{1/3}}{2D} \xi_0  \right) \text{Ai}\left(\frac{2D}{v c_1^{1/3}} + \xi_0 \right),
\end{equation}
\begin{equation}
	\int^{\infty}_{\xi_0} d\xi \; \exp \left( \frac{\xi}{\eta}  \right) \text{Ai} \left(\eta + \xi \right) =
	\frac{N}{n_c c_1^{1/3}} \exp \left( \frac{\xi_0}{\eta}   \right) \text{Ai}\left(\eta + \xi_0 \right),
\end{equation}
where in the second equation we just substituted $\eta = 2Dc_1^{-1/3}/v$ which is a small quantity since the diffusion coefficient is much smaller than the speed.
The next approximation is to extend the limit of integration from the first zero of the Airy function to $-\infty$, by knowing that in this domain the function is oscillating around $0$ and therefore is expected to give a negligible contribution.
This allows us to use the following equality involving the Airy function \cite{widder1979airy}
\begin{equation}
	\int_{-\infty}^\infty e^{p t} \text{Ai}(t) dt = \exp \left( p^{3}/3 \right)
\end{equation}
which leads to the following expression if we neglect $\eta$ in the Airy function argument
\begin{equation}
	\exp \left( \eta^{-3}/3 \right) = \frac{N c_1^{-1/3}}{n_c} \exp \left( \frac{\xi_0}{\eta}   \right) \text{Ai}\left(\eta + \xi_0 \right).
\end{equation}
The next steps is to take the logarithm of this expression and expand the Airy function around its zero
\begin{equation}
	\frac{\eta^{-3}}{3} = \log \left(\frac{N c_1^{-1/3}}{n_c} \right) + \xi_0 \eta^{-1} + \log\left( \eta \text{Ai}'\left(\xi_0\right) \right) .
\end{equation}
We now consider the leading term $\eta^{-3}$ and the logarithmic term containing the population size. 
This gives us the following estimate of the speed in the linear fitness regime (shown in the main text)
\begin{equation}
	v \approx 2 \left( 3 s D^2 \ln\left(\frac{N}{n_c} \frac{s^{1/3}}{D^{1/3}} \right)\right)^{1/3}.
	\label{eq:SM_speed_lin_fitn1}
\end{equation}
{\VC At this stage we can see that for a large enough population size $N \gg n_c (D/s)^{1/3}$ the third term on the right-hand-side of Eq.~\ref{eq:SM_fitness_speed_rel_gen} is negligeable with respect to the second term $\xi_0 (Ds^2)^{1/3}$ and Eq.~\ref{eq:SM_fitness_speed_rel} is a good approximation to the fitness speed relation. This condition is verified for the range of parameters considered in this paper as shown in Fig.~\ref{fig:SM_speed_stoch}A.}

To get a more precise estimate of the speed, one can consider also the order $\eta^{-1}$, leading to
\begin{equation}
	v = 2 \left( s D^2 \right)^{1/3} \left[ \left(3 \ln\left(\frac{N}{n_c} \frac{s^{1/3}}{D^{1/3}} \right)\right)^{1/3} + \xi_0 \left(3 \ln\left(\frac{N}{n_c} \frac{s^{1/3}}{D^{1/3}} \right)\right)^{-1/3}\right].
	\label{eq:SM_linear_fitness_speed}
\end{equation}

In fact, these expressions still depends on $N$ and $s$ which we derive in the following.
We start by integrating the equation for the density of immune receptors, Eq.~2, with a stationary number of infected hosts $N(t) = N$ and defining $\tau = M N_h / N$:
\begin{equation}
h(x, t) = \frac{1}{N} \int_{-\infty}^t \frac{dt'}{\tau} \exp \left( - \frac{t - t'}{\tau} \right) n(x, t) .
\end{equation}
Then one makes the approximation that the wave has a very small width compared with the spatial scale $v\tau$, which characterizes the decay of the immune density.
This allows us to consider $n(x)$ as a delta function at $u = 0$: $n(u) = \delta(u)/N$, where $u = x - vt$, leading to the following expression
\begin{equation}
h(u) = \frac{1}{v \tau} \exp \left( \frac{u}{v \tau} \right) \Theta(-u) ,
\end{equation}
where $\Theta(u)$ is the Heaviside function.
With such an expression, the coverage, Eq.~3, can be computed explicitly, leading  to (for $u>0$)
\begin{equation}
c(u) = \frac{r_0 e^{-u/r_0}}{v \tau + r_0} .
\end{equation}
We can then compute the fitness felt by the wave
\begin{equation}
F(u) = \beta \left(1 - \frac{r_0 e^{-u/r_0}}{v \tau + r_0}\right)^M - \alpha - \gamma .
\end{equation}
In this regime the fitness is assumed to be linear, $F(u) \approx F(0) + \partial_u F(0) u$, which is justified if the width of the wave is much smaller than $r_0$.
By having assumed the stationary condition, we expect the fitness of the bulk of the wave to be zero, i.e. $F(0) = 0$.
The leads to the following condition that connects the wave speed with the population size $N$ and, together with Eq.~\ref{eq:SM_linear_fitness_speed}, closes the system having $v$ and $N$ as unknown
\begin{equation}
v \tau = \frac{v M N_h}{N} = r_0 \left( R_0^{1/M} - 1 \right)^{-1} .
\label{eq:SM_vtau}
\end{equation}

Finally, the explicit value of the fitness slope $s$ can be obtained 
from $s = \partial_u F(0)$
\begin{equation}
s = \frac{\beta M}{r_0 + v \tau} \left( \frac{v \tau}{r_0 + v \tau} \right)^{M-1} = \frac{\alpha + \gamma}{r_0} M \left(R_0^{1/M} - 1\right) .
\label{eq:SM_linear_fitness_s}
\end{equation}

\section{Derivation of the mutational load}
\label{sec:mut_load}

To obtain the effect of deleterious mutations in our epidemiological context, we follow the approach proposed in \cite{koelle2015effects}, which, in turn, refers to the classical results of mutation selection balance of population genetics \cite{kimura1966mutational, haigh1978accumulation}.
We consider a population in which $n_k(t)$ is the number of individuals carrying $k$ deleterious mutations.
Mutations are assumed to occur during the bottleneck of a transmission event.
This is because harmful mutations arising within the very few individuals that are transmitted are weakly subject to purifying selection, while, if they occur during the course of an in-host infection, selection will tend to remove them.
The number of mutations that can  occur per genome at transmission are assumed to follow a Poisson distribution with rate $U_d$.
Moreover, we assume that each single deleterious mutation affects the transmissibility of the population by a multiplicative factor $(1-s_d)$, leading to the following transmissibility for the population having $k$ mutations:
\begin{equation}
\beta_k = \beta_0 (1 - s_d)^k .
\end{equation}

Putting all the assumptions together, one obtain the following temporal evolution for the number of infected hosts:
\begin{equation}
\partial_t n_k = S \sum_{j=0}^k \beta_j n_j Poiss(k-j|U_d) - \gamma n_k = S \beta_0 e^{-U_d} \sum_{j=0}^k (1 - s_d)^j n_j  \frac{U_d^{k-j}}{(k-j)!} - \gamma n_k
\label{eq:SM_n_mut_k}
\end{equation}
where, for simplicity, the virulence $\alpha$ and the recovery rate $\gamma$ are condensed together in a single parameter.

At equilibrium, one can impose the stationarity of the equation above and find the number of infected hosts
\begin{equation}
n_k^* = N \frac{e^{-U_d/s_d}}{k!}  \left (\frac{U_d}{s_d} \right)^k .
\end{equation}
This expression can be verified by substituting it in Eq.~\ref{eq:SM_n_mut_k} and using the relation $\beta_0 S = e^{U_d} \gamma$ that can be obtained from $\partial_t n_0 = 0$.

As a final step, we can compute the average transmission rate that such a population has
\begin{equation}
\langle \beta \rangle = \sum_k \beta_k \frac{n^*_k}{N} =
\beta_0 e^{-U_d/s_d} \sum_k \frac{(1-s_d)^k}{k!}  \left (\frac{U_d}{s_d} \right)^k =
\beta_0 e^{-U_d} \approx \beta_0 (1 - U_d).
\end{equation}
Using this expression, one can then obtain an effective growth rate for the population having a given deleterious mutation rate
\begin{equation}
F(x) = \langle \beta \rangle S(x) - \gamma = \beta_0 S(x) e^{-U_d} - \gamma \approx \beta_0 S(x) (1 - U_d) - \gamma .
\label{eq:SM_mutation_tradeoff}
\end{equation}

In the main text, to discuss about evolutionary stability of the beneficial mutation rate or selection coefficient, the deleterious mutation coefficient is expressed as the product of a constant and the beneficial mutation rate
\begin{equation}
U_d = a \mu_x = \lambda D
\end{equation}
where $a = \Delta x^2 \lambda/2$ can be interpreted as a ratio between deleterious and beneficial mutations which cannot be changed.
What can be changed by viruses is the global mutation rate, which would increase the antigenic mutation rate $\mu_x$ but, at the same time, would increase the deleterious rate though the relation above, leading to the mutational trade-off.

\section{Evolutionary stability analysis}

\subsection{General stability condition}\label{sec:evo_stab}

To derive the condition for the invasion of a mutant, we start by considering a resident population at a stationary-wave state.
We also consider a generic mutant that can have, in general, a new set of parameters labeled with a prime, e.g. $\beta'$, $D'$, $\ldots$, which are assumed to be close to the parameters of the resident.
In the frame of reference of the resident wave moving at speed $v$, the equation for the mutant dynamics reads
\begin{equation}
\begin{aligned}
&\partial_t n'(u,t) = D' \partial_u^2 n'(u,t) + v \partial_u n'(u,t) + F'(u) \Theta(n+n' > n_c) n'(u)\\
&F'(u) = \beta' S(u) - \gamma' -\alpha' \approx F'_T + s'_T (u - u_T).
\end{aligned}
\end{equation}
Note that the quantities not labeled with a prime are the wave speed $v$ and the susceptibility of the resident population $S(u)$.
We assume that the mutant is rare enough that it does not generate any significant immune response, and, therefore, it does not contribute to the susceptibility.
Moreover, the resident population number appears within the theta function, which imposes the cutoff when the total number of individuals, $n+n'$, is smaller than the threshold $n_c$.
This assumption allows us to identify the tip of the wave at the same position both for the resident and the invading populations, greatly simplifying the calculations.
Finally, as for the previous calculations, the fitness is linearized around the tip, implying that, also for this derivation, the success or failure of an invasion depends only on what happens at the tip.

\begin{figure*}
	\includegraphics[width=0.95\textwidth]{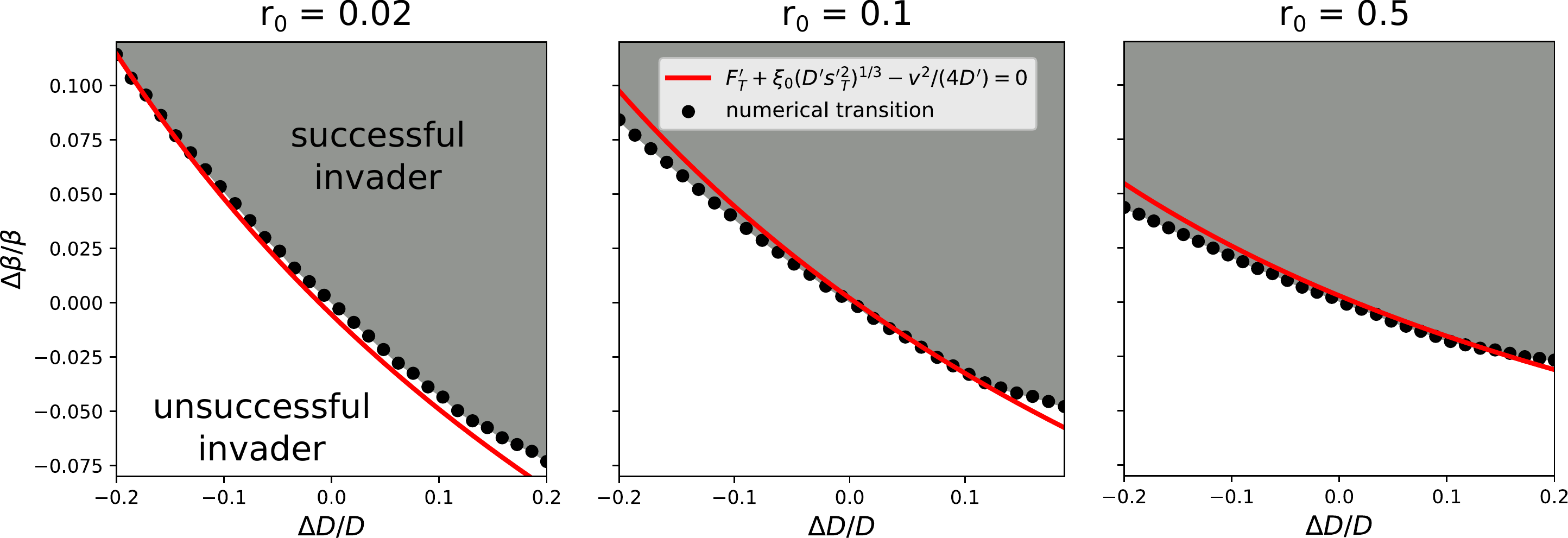}
	\caption{Testing the invasion criteria \ref{eq:SM_invasion_eq}.
		A given resident population is first simulated in isolation until it reaches stationarity with parameters $D=5 \cdot 10^{-6}$, $\beta=2$, $\gamma+\alpha=1$, $M=5$, $N_h=10^{12}$ and $r_0$ is indicated in the plot title.
		Then, a mutant with given $D' = D + \Delta D$ and $\beta' = \beta + \Delta \beta$ is introduced with $n'(x) = \epsilon n(x)$, $\epsilon = 0.05$ (all the other mutant parameters are the same of the resident).
		The system evolves until one of the two populations becomes $10/\epsilon$ times bigger than the other or after $6000$ units of time.
		This is repeated for different values of $\beta'$ using a bisection-like iteration  until the point of transition, $\tilde{\beta}$, between a successful or unsuccessful invader is found.
		Each black point in the plot is $(\beta^*-\beta)/\beta$ for a given $D'$.
		The red line is the prediction of Eq.~\ref{eq:SM_invasion_eq} equal to zero.
	}
	\label{fig:SM_inv_diagram}
\end{figure*}

We are going to look for solutions $n'(u,t) = e^{\rev{\rho} t} \phi(u)$, i.e. a stationary profile that would grow or decay at rate $\rev{\rho}$.
Here we also make the approximation that success or failure in the invasion depends only on the sign of this \rev{rho}.
That is to say that we identify a successful mutant only by looking at its initial growth rate.
By substituting this solution in the equation above with a linearized tip we can solve the equation on the right and on the left of the tip as performed in the previous paragraphs.
For $u > u_T$ one has to solve
\begin{equation}
D' \partial_u^2 \phi(u) + v \partial_u \phi(u) - \rev{\rho} \phi(u) = 0 ,
\end{equation}
which leads to the solution
\begin{equation}
\phi(u) =  n_c \exp\left(- \frac{v}{2 D'} \left(1 + \sqrt{1 + \frac{4 D'}{v^2}\rev{\rho} } \right) (u-u_T) \right) \approx n_c \exp\left(-\left( \frac{v}{D'} + \frac{\rev{\rho}}{v} \right) (u-u_T) \right)
\end{equation}
On the left side of the tip, we can find an Airy equation like Eq.~\ref{eq:SM_Airy} (but different coefficients $c_1$ and $c_2$)
\begin{equation}
\frac{D'}{s'_T} \partial^2_u \phi(u) + \left(\frac{F'_T - \rev{\rho}}{s'_T} - u_T - \frac{v^2}{4 D' s'_T} + u \right) \phi(u) \equiv c_1 \partial^2_u \phi(u) + \left(u - c_2\right) \phi(u) = 0 .
\end{equation}
Therefore leading to the solution \ref{eq:SM_n_left}.
As before we impose the continuity of the function and the derivative at the intersection, leading to
\begin{equation}
\frac{\text{Ai}'\left( \xi_0 + \epsilon \right)}{\text{Ai}\left( \xi_0 + \epsilon \right)} = \left( \frac{v }{2 D} + \frac{\rev{\rho}}{v}\right) c_1^{1/3} \approx \frac{v }{2 D} c_1^{1/3}.
\end{equation}
where $\rev{\rho}$ is considered to be small.
We can then carry out all the procedure of the sections before of approximating the Airy function around its zero.
This leads to
\begin{equation}
c_2 - u_T - \xi_0 c_1^{1/3} = \frac{\rev{\rho} - F_T'}{s_T'} + \frac{v^2}{4 D' s_T'} - \xi_0 \left( \frac{D'}{s_T'} \right)^{1/3}  = 0 ,
\end{equation}
\begin{equation}
\rev{\rho} = F_T' + \xi_0 \left( D' {s_T'}^2 \right)^{1/3} -\frac{v^2}{4 D'} = 0 .
\label{eq:SM_invasion_eq}
\end{equation}
If this last expression is larger than zero, we then expect a mutant that grows and invades the resident population, Eq.~10 of the main text.
This condition has been tested in figure \ref{fig:SM_inv_diagram}, where, given a mutation coefficient $D'$ for the mutant, we looked for the value of transmissibility $\tilde{\beta}'$ such that the mutant invades for $\beta' > \tilde{\beta}'$ or does not for $\beta' < \tilde{\beta}'$.
The equation above, i.e. $\rev{\rho}(\tilde{\beta}', D') = 0$, provides a prediction for this value $\tilde{\beta}'$ as a function of $D'$.
Despite the numerous approximations in the computation above, the prediction of this transition point is very accurate.
More details on how the simulations are performed are in the caption of the figure.

The invasion condition \ref{eq:SM_invasion_eq} simplifies considerably in the
limits of small and large $r_0$. For small $r_0$, the fitness is
saturated, so that $F'_T=F'_{\rm max}=\beta' -\alpha'-\gamma'$ and $\sigma'_T=0$. Using
$F'_{\rm max}=v'^2/4D'$, the invasion condition becomes:
\beq
\frac{v'^2-v^2}{4D} \rev{>0}.
\eeq
The evolutionary stable solution is the one that maximizes the speed
of the wave.

For large $r_0$ the fitness profile is approximately linear, so that $s_T' = s'$ and $F_T' = F(0)' + s' x_T$, $F_T = s x_T$.
By plugging these equations into the invasion condition one has
\begin{equation}
F(0)' + s' x_T + \xi_0 \left( D' s'^2 \right)^{1/3} - \frac{v^2}{4 D'} > 0 .
\end{equation}
We can now use the speed-fitness relation, Eq.~7, and $F_T = s x_T$ to obtain
\begin{equation}
F(0)' + x_T \left(s' - \frac{D}{D'} s\right) +
\xi_0  \left( \left( D' s'^2 \right)^{1/3} - \frac{D}{D'} \left( D s^2 \right)^{1/3} \right) > 0 .
\end{equation}
The selection terms in $s$ and $s'$ are subdominant, since $s \propto r_0^{-1}$.
The dominant term is therefore the fitness of the mutant at the center of the wave,
\begin{equation}
F(0)' = \beta' S(0) - \alpha' - \gamma' = \frac{\beta'(\alpha + \gamma)}{\beta} - \alpha' - \gamma' >0.
\end{equation}
Using the definition of the reproductive ratio $R_0 = \beta /(\alpha +
\gamma)$, this yields the condition
\begin{equation}
R_0' > R_0 .
\end{equation}

To find a general criterion for the evolutionary stability of the viral population, we assume that the evolution acts on a generic parameter $\theta$ from which all the other parameters can depend on: $\beta(\theta)$, $\alpha(\theta)$, $\gamma(\theta)$, $D(\theta)$.
As before, we label with a prime the parameter of a mutant $\theta'$.
We also indicate the fitness and its derivative at the tip as $F_T(\theta)$, $s_T(\theta)$ for the resident population and $F_T'(\theta', \theta) = \beta(\theta')S(\theta) - \alpha(\theta') - \gamma(\theta')$, $s'_T(\theta', \theta) = \beta(\theta') \partial_x S(\theta)$ for the mutant growing in the resident $\theta$.
The growth rate of a mutant can be then expressed as a function of $\theta$ and $\theta'$: $\rev{\rho}(\theta', \theta)$.
The evolutionary stability is reached at a value $\theta^*$ such that the growth rate of a mutant having a slightly different value is no larger.
As shown in the main text, this translates into the condition $\partial_{\theta'} \rev{\rho}(\theta', \theta) |_{\theta'=\theta=\theta^*} = 0$, where the derivative acts only on the parameters labeled with a prime in equation \ref{eq:SM_invasion_eq},
\begin{equation}
\left. \left[ \partial_{\theta'} F_T'(\theta', \theta) + \frac{2 \xi_0}{3} \left(\frac{D(\theta')}{s_T'(\theta', \theta)}\right)^{1/3} \partial_{\theta'} s_T'(\theta', \theta) + \left(\frac{v(\theta)^2}{4 D(\theta')^2} + \frac{\xi_0}{3} \left(\frac{s_T'(\theta', \theta)}{D(\theta')}\right)^{2/3} \right) \partial_{\theta'} D(\theta') \right] \right|_{\theta'=\theta=\theta^*}= 0 .
\end{equation}
We now want to express the equation only as a function of the fitness and the selection at the tip by using Eq.~\ref{eq:SM_fitness_speed_rel} for $v(\theta)$.
We will also use the fact that $D(\theta)|_{\theta = \theta^*} = D(\theta')|_{\theta' = \theta^*}$
\begin{equation}
\left. \left[ \partial_{\theta'} F_T'(\theta', \theta) + \frac{2 \xi_0}{3} \left(\frac{D(\theta')}{s_T'(\theta', \theta)}\right)^{1/3} \partial_{\theta'} s_T'(\theta', \theta) + \left(F_T'(\theta', \theta) + \frac{4 \xi_0}{3} \left({s_T'(\theta', \theta)}^2D(\theta')\right)^{1/3} \right) \frac{\partial_{\theta'} D(\theta')}{D(\theta')}\right]\right|_{\theta'=\theta=\theta^*} = 0 .
\label{eq:SM_ES_theta}
\end{equation}
This expression can be rewritten in a more compact form by introducing $\sigma_T'(\theta', \theta) = \xi_0(D(\theta') s_T'(\theta', \theta)^2)^{1/3}$, which leads to
\begin{equation}
\left. \left[ \frac{\partial_{\theta'} F_T'(\theta', \theta) + \partial_{\theta'} \sigma_T'(\theta', \theta)}{F_T'(\theta', \theta) + \sigma_T'(\theta', \theta)} + \frac{\partial_{\theta'} D(\theta')}{D(\theta')} \right]\right|_{\theta'=\theta=\theta^*} = 0 .
\label{eq:SM_ES}
\end{equation}
Recognizing logarithmic derivatives, this condition is equivalent to:
\begin{equation}
\left. \partial_{\theta'}\left[\left(F_T'(\theta', \theta) +\sigma_T'(\theta', \theta) \right)D(\theta')\right]\right|_{\theta'=\theta=\theta^*}=0.
\end{equation}

\subsection{Evolutionary stability of the mutation rate}
\label{sec:evo_stab_mutrate}

\begin{figure*}
	\includegraphics[width=\textwidth]{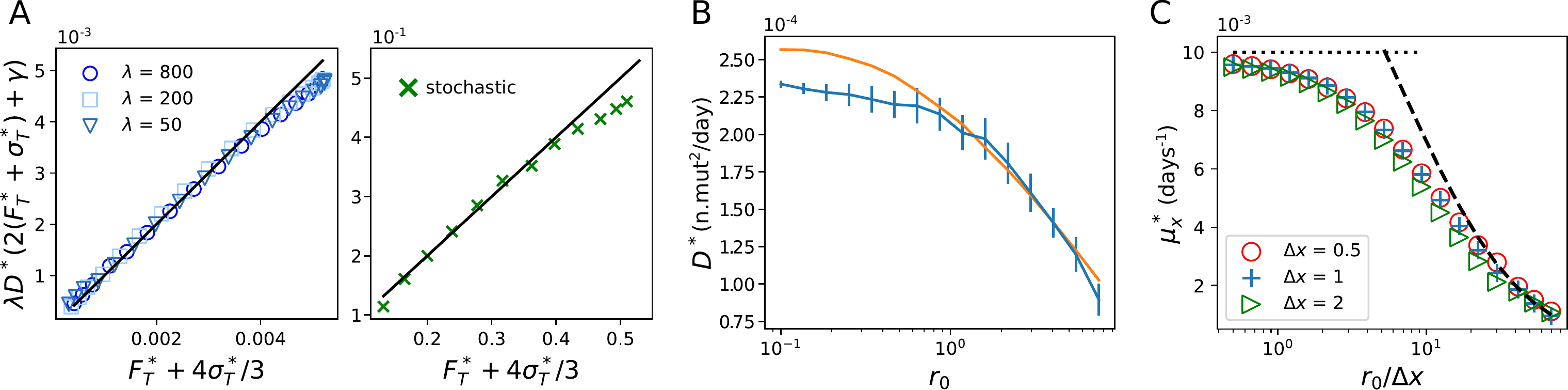}
	\caption{
	\rev{Panel (A): test of the evolutionary stability equation Eq.~\ref{eq:SM_ES_mut_rate}.
	The left panel shows the deterministic setting with cutoff with parameters $\beta_0=0.05~\text{days}^{-1}$, $\gamma+\alpha=0.04~\text{days}^{-1}$, $M=5$, $N_h=10^{10}$.
	The right panel A tests the equation for the stochastic simulations as described in Sec.~4B with $n_{stoch} = 10^4$ and parameters $\lambda=250~\text{days/n.mutations}^2$, $\beta=2~\text{days}^{-1}$, $\gamma+\alpha=1~\text{days}^{-1}$, $M=5$, $N_h=10^{10}$.
	(B): same stochastic simulations of panel A-right but plotted as a function of $r_0$.
	The blue line is the temporal average over $D^*$ which fluctuates with a the standard deviation of the error bar.
	The continuous line is the prediction of Eq.~\ref{eq:SM_ES_mut_rate}.
	Panel (C) checks that the evolutionary stable antigenic mutation rate is independent of $\Delta x$ after a proper rescaling of $r_0$. This confirms that Eq.~5 is approximately invariant by spatial re-scaling. 
	Parameters: $\beta_0=0.05~\text{days}^{-1}$, $\gamma+\alpha=0.04~\text{days}^{-1}$, $a = 100$ days, $M=5$, $N_h=10^{10}$.}
	}
	\label{fig:SM_es_mut_rate_stoch}
\end{figure*}

The evolutionary stable antigenic diffusion coefficient $D^*$ under mutational load trade-off, where the fitness is $F(u) = \beta_0 S(u)(1 - \lambda D) - \gamma - \alpha$, from Eq.~\ref{eq:SM_mutation_tradeoff}, can be obtained by identifying $\theta = D$ in Eq.~\ref{eq:SM_ES}.
Note that the choice $\Delta x = 1$ allows us to get the evolutionary stable antigenic mutation rate as $U_x^* = 2 D^*$.
After some algebra, one can get the following formula
\begin{equation}
F_T^* (1 - 2 \lambda D^*) + \sigma_T^* \left( \frac{4}{3} - 2{\lambda} D^*  \right) - \gamma \lambda D^* = 0 ,
\label{eq:SM_ES_mut_rate}
\end{equation}
where for simplicity we put $\alpha = 0$ and we dropped the dependencies from $\theta^*$, writing, for example, $F_T^* = F_T(\theta^*)$.
This expression provides $D^*$ as a function of the fitness value and slope at the tip and it is tested in Fig.~3 and \ref{fig:SM_es_mut_rate_stoch}, that prove its validity also in the stochastic setting.

It is interesting to study the limits of this expression in the FKPP and linear fitness regime.
In the first, setting $\sigma_T = 0$ and $F_T = \beta_0(1- \lambda D) - \gamma$, one can obtain
\begin{equation}
D^* = \frac{\beta_0 - \gamma}{2 \beta_0 \lambda},
\label{eq:SM_ES_mut_rate_FKPP}
\end{equation}
under the condition that $1 - \lambda D = 1 - U_d \neq 0$, which is satisfied since the deleterious mutation rate is a small quantity.

In the linear fitness regime, an explicit expression can be obtained only by considering $\lambda D \ll 1$, which leads to
\begin{equation}
F_T^* + \sigma_T^* \frac{4}{3} - \gamma {\lambda} D^* = \frac{{v^*}^2}{4 D^*} + \frac{\sigma_T^*}{3} - \gamma {\lambda} D^* = 0 ,
\end{equation}
where we used the fitness-speed relation.
We can now express the speed as $v = A \; s^{1/3} D^{2/3}$ using Eq.~\ref{eq:SM_linear_fitness_speed}, where $A$ contains logarithmic dependencies.
This allows us to make also the approximation of considering $A$ constant in $D$ and get
\begin{equation}
D^* = \frac{M \gamma}{r_0} \left(\left(\frac{\beta_0}{\gamma}\right)^{1/M} - 1\right) \left( \frac{1}{\gamma {\lambda}} \left( \frac{A^2}{4} - \frac{\xi_0}{3}\right) \right)^{3/2} ,
\end{equation}
where we used Eq.~\ref{eq:SM_linear_fitness_s} for the selection coefficient.

\subsection{Evolutionary stability of the virulence}
\label{sec:evo_stab_virulence}

\begin{figure*}
	\includegraphics[width=0.95\textwidth]{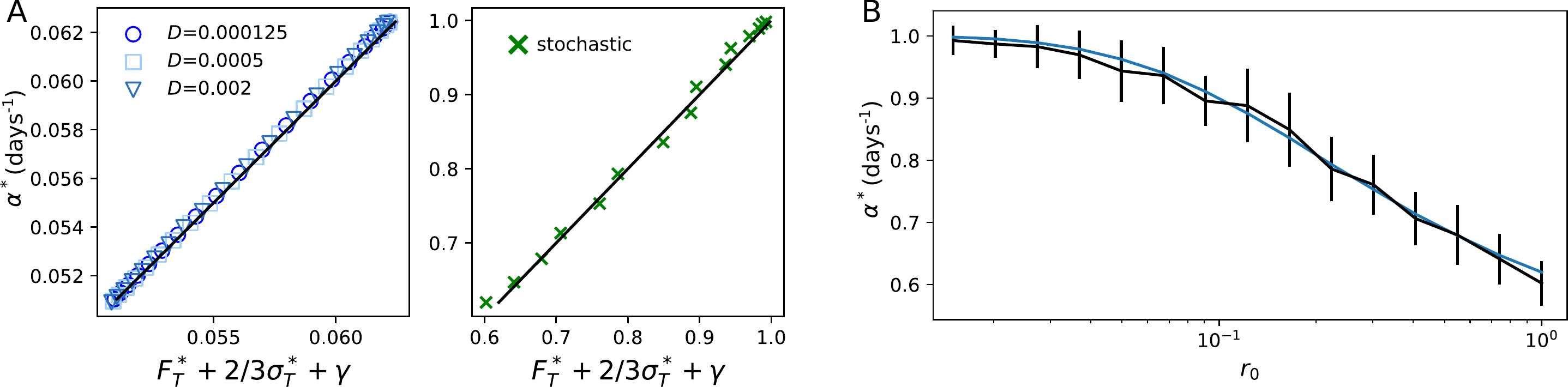}
	\caption{ \rev{Panel (A):  test of Eq.~\ref{eq:SM_ES_virulence}.
		On the left in the deterministic setting with cutoff, where the transmissibility reads $\beta(\alpha) = b \sqrt{\alpha}$ and the  parameters are $b=0.5~\text{days}^{-1/2}$, $\gamma=0.05~\text{days}^{-1}$, $M=5$, $N_h=10^{10}$.
		On the right in the stochastic setting with parameters: $D=10^{-5}~\text{n.mutations}^{2}\text{/days}$, $b = 2~\text{days}^{-1/2}$, $\gamma = 0.5~\text{days}^{-1}$, $M=5$, $N_h = 10^{12}$.
		The simulations are performed as described in Sec.~3B and Sec.~3C with $ n_{stoch} = 10^5$.
		The points are temporal averages of the quantity.
		Panel (B): same stochastic simulation shown as a function of $r_0$.
		The error bars quantifies the standard deviations of the temporal fluctuations of $\alpha^*$.
		The blue line is the prediction of Eq.~\ref{eq:SM_ES_virulence}. }
	}
	\label{fig:SM_es_virulence_stoch}
\end{figure*}

In a similar way of what we did for the evolutionary stable mutation rate, we can obtain the equation for the evolutionary stable virulence, i.e. $\theta = \alpha$ in Eq.~\ref{eq:SM_ES},
\begin{equation}
\frac{\partial_\alpha \beta(\alpha^*)}{\beta(\alpha^*)} \left(F_T^* + \alpha^* + \gamma + \frac{2}{3}\sigma_T^* \right) = 1, \hspace{0.5cm} \alpha^* = F_T^*  + \gamma + \frac{2}{3}\sigma_T^* ,
\label{eq:SM_ES_virulence}
\end{equation}
where the expression on the right assumes the transmissibility trade-off as $\beta(\alpha) = b \sqrt{\alpha}$.
The validity of this expression is tested in Fig.~5 in a deterministic setting and Fig.~\ref{fig:SM_es_virulence_stoch} for a stochastic simulation.

We can also write explicitly the expression in the FKPP regime (in a general way and with our specific assumption on the trade-off):
\begin{equation}
\partial_\alpha \beta(\alpha^*) = 1, \hspace{0.5cm} \alpha^* = b^2/4 .
\label{eq:SM_ES_virulence_FKPP}
\end{equation}

Finally, in the linear fitness regime one can simply use the fitness-speed relation to express $F_T$ as a function of the speed and get
\begin{equation}
\frac{\partial_\alpha \beta(\alpha^*)}{\beta(\alpha^*)} \left(\frac{{v^*}^2}{4 D^*} + \alpha + \gamma - \frac{\sigma_T^*}{3} \right) = 1.
\label{eq:SM_ES_virulence_lin}
\end{equation}
One can then express $v$ from Eq.~\ref{eq:SM_linear_fitness_speed}, use $s_T^*$ given by Eq.~\ref{eq:SM_linear_fitness_s} and numerically solve for $\alpha^*$ (which is the procedure used to get the linear-fitness predictions in -Fig.5.)

\end{document}